\documentclass[12pt]{article}  
\usepackage[a4paper,textwidth=490.0pt,textheight=703.1pt]{geometry}
\usepackage{slashed}
\usepackage{graphicx}
\usepackage{epsfig}
\usepackage{amsmath}
\usepackage{amssymb}
\usepackage{ulem}
\usepackage{mdwlist}
\usepackage{color}                                                  
\usepackage{float}
\usepackage[rflt]{floatflt}
\usepackage{slashed}          
\usepackage{cite}
\usepackage{mathtools}
\usepackage{dsfont}
\usepackage{subfig}
\newcommand{\E}{\mathrm{e}}
\newcommand{\II}{\mathrm{i}}
\newcommand{\T}{\mathsf{T}}
\newcommand{\EulerGamma}{\gamma_\text{E}}
\newcommand{\rmd}{\mathrm{d}}

\newcommand{\N}{\nonumber}

\usepackage{array}

\usepackage[english]{babel}
\usepackage[latin1]{inputenc}
\usepackage[T1]{fontenc}
\usepackage{ae}

\usepackage{url}

\usepackage{amsmath, amsthm, amssymb}

\newtheorem{thm}{Theorem}[section]

\usepackage{array}

\usepackage[english]{babel}
\usepackage[latin1]{inputenc}
\usepackage[T1]{fontenc}
\usepackage{ae}

\newtheorem{definition}[thm]{Definition}

\newcommand{\ep}{\varepsilon}

\usepackage{rotating}

\usepackage{graphicx}

\newcounter{mmacnt}
\def\restartmma{\setcounter{mmacnt}{0}}
\restartmma \catcode`|=\active
\def|#1|{\mathrm{#1}}
\catcode`|=12
\newenvironment{mma}{
 \par\smallskip
 \catcode`|=\active
 \parskip=0pt\parindent=0pt 
 \small
 \def\In##1\\{%
   \def\linebreak{\hfill\break\null\qquad}%
   \refstepcounter{mmacnt}
   \hangindent=2.5em\hangafter=0
   \leavevmode
   \llap{\tiny\sffamily In[\arabic{mmacnt}]:=\kern.5em}%
   \mathversion{bold}\footnotesize$\displaystyle##1$\normalsize
   \mathversion{normal}\par
 }%
 \def\Print##1\\{%
   \def\linebreak{\hfill\break}%
   \hangindent=2.5em\hangafter=0
   \leavevmode ##1\par}%
 \def\Out##1\\{%
   \def\linebreak{$\hfill\break\null\hfill$}%
   \kern\abovedisplayskip\par
   \hangindent=2.5em\hangafter=0
   \leavevmode
   \llap{\tiny\sffamily Out[\arabic{mmacnt}]=\kern.5em}
   \footnotesize$\displaystyle##1$\normalsize\hfill\null\par
   \kern\belowdisplayskip
 }%
 \def\Warning##1##2\\{%
   \def\linebreak{\hfill\break}%
   \hangindent=2.5em\hangafter=0
   \leavevmode
   {\scriptsize##1 : ##2}\par}%
}{%
 \par\smallskip
}

\usepackage{color}

\newenvironment{fshaded}{%
\MakeFramed {\FrameRestore}
}%
{\endMakeFramed}

\usepackage{tikz}
\usetikzlibrary{matrix}

\allowdisplaybreaks[4]

\begin{document}
\setlength{\baselineskip}{0.515cm}
\sloppy
\thispagestyle{empty}
\begin{flushleft}
DESY 17--062
\hfill  
\\
DO--TH 17/05  \\
MSUHEP-17-004 \\
MITP/17-033   \\
TTK-17-13     \\
April 2017\\
\end{flushleft}

\setcounter{table}{0}

\mbox{}
\vspace*{\fill}
\begin{center}

{\Large\bf \boldmath The Three-Loop Splitting Functions $P_{qg}^{(2)}$ and $P_{gg}^{(2,\rm  N_F)}$} 

\vspace{4cm}
\large
J.~Ablinger$^a$,
A.~Behring$^{b,c}$,
J.~Bl\"umlein$^b$,
A.~De Freitas$^b$,
A.~von Manteuffel$^d$,

\vspace*{1mm}
and C.~Schneider$^a$

\vspace{1.5cm}
\normalsize   

{\it $^a$~Research Institute for Symbolic Computation (RISC),
                          Johannes Kepler University, Altenbergerstra\ss{}e 69,
                          A--4040, Linz, Austria}

\vspace*{2mm}
{\it $^b$~Deutsches Elektronen-Synchrotron, DESY, Platanenallee 6, D-15738 Zeuthen, Germany}

\vspace*{2mm}
{\it $^c$~Institut f\"ur Theoretische Teilchenphysik und Kosmologie,
 RWTH Aachen University, D--52056 Aachen, Germany}

\vspace*{2mm}
{\it $^d$~Department of Physics and Astronomy, Michigan State University, \\ East Lansing, MI 48824, 
USA}
\\

\end{center}
\normalsize
\vspace{\fill}
\begin{abstract}
\noindent 
We calculate the unpolarized twist-2 three-loop splitting functions $P_{qg}^{(2)}(x)$ and 
$P_{gg}^{(2,\rm N_F)}(x)$ and the associated anomalous dimensions using massive three-loop 
operator matrix elements. While we calculate $P_{gg}^{(2,\rm N_F)}(x)$ directly, $P_{qg}^{(2)}(x)$ 
is computed from 1200 even moments, without any structural prejudice, using a hierarchy of 
recurrences obtained for the corresponding operator matrix element. The largest recurrence 
to be solved is of order 12 and degree 191. We confirm results in the foregoing literature.
\end{abstract}

\vspace*{\fill}
\noindent
\numberwithin{equation}{section}

\newpage
\section{Introduction}
\label{sec:1}

\vspace*{1mm}
\noindent
The three-loop anomalous dimensions and splitting functions $P_{ij}^{(2)}(x)$ govern the 
scale-evolution of the parton distribution functions in Quantum Chromodynamics (QCD) at next-to-next-to-leading 
order (NNLO) 
and are of importance for precision predictions at $ep$- and hadron colliders, such as HERA, 
the Tevatron and the LHC. They are instrumental for the precise prediction of the Higgs-, top-quark,
and jet-production cross sections, concerning both, the parton distribution functions 
\cite{Accardi:2016ndt,Alekhin:2017kpj} and the strong coupling constant $\alpha_s(M_Z)$ 
\cite{alphas}, thus allowing the exploration of the hitherto heaviest systems in the Standard 
Model in greatest possible detail. 
High precision predictions are also of importance for planned $ep$-facilities as EIC and LHeC \cite{FUTURE}, 
operating at high luminosity. A first computation of the three-loop splitting functions was performed in 
Refs.~\cite{Moch:2004pa,Vogt:2004mw}, after the calculation of fixed moments for these 
quantities in Refs.~\cite{MOMA,Blumlein:2009rg,Bierenbaum:2009mv}. In these references the anomalous 
dimensions were calculated using massless processes.

In the present paper we compute the splitting function $P_{qg}^{(2)}$ and the $N_F$-dependent part of 
$P_{gg}^{(2)}$ independently using massive three-loop operator matrix elements (OMEs). They are necessary 
for the computation of the heavy flavor contributions to deep-inelastic scattering in the region of virtualities 
$Q^2$ much larger than the heavy quark mass squared $m^2$. While we obtain the complete 
expression for $P_{qg}^{(2)}$, only the $N_F$-part of $P_{gg}^{(2)}$ emerges, since the 
respective process is proportional to the color factor $T_F$.\footnote{In $SU(N_c)$ the color factors 
contributing to the three-loop anomalous dimensions are given by $T_F = 1/2, C_A = N_c, C_F = 
(N_c^2-1)/(2N_c)$, with $N_c$ the number of colors and $N_c = 3$ in case of QCD.}

We calculate $P_{gg}^{(2, \rm N_F)}$ directly by applying the techniques described in 
Ref.~\cite{Ablinger:2015tua}. In the case of $P_{qg}^{(2)}$, {we apply} the 
recently proposed method of large Mellin moments \cite{Blumlein:2017dxp}. In previous papers 
we have obtained the corresponding contributions $\propto T_F$ for the splitting functions
$P_{gq}^{(2, \rm N_F)}, P_{qq}^{(2, \rm NS, N_F)}, P_{qq}^{(2, \rm NS, TR, N_F)}$ 
\cite{Ablinger:2014lka,Ablinger:2014vwa} 
and the complete splitting function $P_{qq}^{(2), \rm PS}$ \cite{Ablinger:2014nga}. A series 
of Mellin moments for these quantities has been obtained before in Ref.~\cite{Bierenbaum:2009mv}. 
Our results agree with those of the previous calculation in Refs.~\cite{Moch:2004pa,Vogt:2004mw}.
{Recently, parts} of the three-loop anomalous dimensions $\propto T_F$ have also been 
obtained for a series of anomalous dimensions in two-mass calculations \cite{TWOMASS}.

The paper is organized as follows. In Section~2 we describe the calculation of the anomalous 
dimensions $\gamma_{qg}^{(2)}$ and $\gamma_{gg}^{(2),\rm N_F}$ in Mellin-$N$ space.  Section~3 contains the 
conclusions. In the Appendices we present the splitting functions in $x$-space and give some technical 
details in calculating master integrals. In all representations we project onto the algebraic basis 
\cite{Blumlein:2003gb}, both for 
the harmonic sums  \cite{Vermaseren:1998uu,Blumlein:1998if} and the harmonic polylogarithms 
(HPLs)~\cite{Remiddi:1999ew}.
\section{The Anomalous Dimensions}
\label{sec:2}

\vspace*{1mm}
\noindent
In the following, we will use the massive OMEs $A_{Qg}^{(3)}$ and $A_{gg,Q}^{(3)}$, 
cf.~Ref.~\cite{Bierenbaum:2009mv}, to calculate the anomalous dimensions. 
The unrenormalized OMEs  {$\hat{\hat{A}}_{Qg}^{(3)}$} and   {$\hat{\hat{A}}_{gg,Q}^{(3)}$} were given 
in~\cite{{Bierenbaum:2009mv}}
\begin{eqnarray}
   \hat{\hat{A}}_{Qg}^{(3)}&=&
                  \Bigl(\frac{\hat{m}^2}{\mu^2}\Bigr)^{3\ep/2}
                     \Biggl[
           \frac{\hat{\gamma}_{qg}^{(0)}}{6\ep^3}
             \Biggl(
                   (N_F+1)\gamma_{gq}^{(0)}\hat{\gamma}_{qg}^{(0)}
                 +\gamma_{qq}^{(0)}
                                \Bigl[
                                        \gamma_{qq}^{(0)}
                                      -2\gamma_{gg}^{(0)}
                                      -6\beta_0
                                      -8\beta_{0,Q}
                                \Bigr]
                 +8\beta_0^2
\N\\ &&
                 +28\beta_{0,Q}\beta_0
                 +24\beta_{0,Q}^2
                  +\gamma_{gg}^{(0)}
                                \Bigl[
                                        \gamma_{gg}^{(0)}
                                       +6\beta_0
                                       +14\beta_{0,Q}
                                \Bigr]
             \Biggr)
          +\frac{1}{6\ep^2}
             \Biggl(
                   \hat{\gamma}_{qg}^{(1)}
                      \Bigl[
                              2\gamma_{qq}^{(0)}
                             -2\gamma_{gg}^{(0)}
                             -8\beta_0
\N\\ &&
                             -10\beta_{0,Q}
                      \Bigr]
                  +\hat{\gamma}_{qg}^{(0)}
                      \Bigl[
                              \hat{\gamma}_{qq}^{(1), {\sf PS}}\{1-2N_F\}
                             +\gamma_{qq}^{(1), {\sf NS}}
                             +\hat{\gamma}_{qq}^{(1), {\sf NS}}
                             +2\hat{\gamma}_{gg}^{(1)}
                             -\gamma_{gg}^{(1)}
                             -2\beta_1
                             -2\beta_{1,Q}
                      \Bigr]
\N\\
&&
                  + 6 \delta m_1^{(-1)} \hat{\gamma}_{qg}^{(0)}
                      \Bigl[
                              \gamma_{gg}^{(0)}
                             -\gamma_{qq}^{(0)}
                             +3\beta_0
                             +5\beta_{0,Q}
                      \Bigr]
             \Biggr)
          +\frac{1}{\ep}
             \Biggl(
                   \frac{\hat{\gamma}_{qg}^{(2)}}{3}
                  -N_F \frac{\hat{\tilde{\gamma}}_{qg}^{(2)}}{3}
                  +\hat{\gamma}_{qg}^{(0)}\Bigl[
                                    a_{gg,Q}^{(2)}
                                   -N_Fa_{Qq}^{(2),{\sf PS}}
                                          \Bigr]
\N\\
&&
                  +a_{Qg}^{(2)}
                      \Bigl[
                              \gamma_{qq}^{(0)}
                             -\gamma_{gg}^{(0)}
                             -4\beta_0
                             -4\beta_{0,Q}
                      \Bigr]
                  +\frac{\hat{\gamma}_{qg}^{(0)}\zeta_2}{16}
                      \Bigl[
                              \gamma_{gg}^{(0)} \Bigl\{
                                                        2\gamma_{qq}^{(0)}
                                                       -\gamma_{gg}^{(0)}
                                                       -6\beta_0
                                                       +2\beta_{0,Q}
                                                \Bigr\}  
\N\\ &&
                             -(N_F+1)\gamma_{gq}^{(0)}\hat{\gamma}_{qg}^{(0)}
                             +\gamma_{qq}^{(0)} \Bigl\{
                                                       -\gamma_{qq}^{(0)}
                                                       +6\beta_0
                                                \Bigr\}
                             -8\beta_0^2
                             +4\beta_{0,Q}\beta_0
                             +24\beta_{0,Q}^2
                      \Bigr]
                  + \frac{\delta m_1^{(-1)}}{2}
                      \Bigl[
                              -2\hat{\gamma}_{qg}^{(1)}
\N\\ &&
                              +3\delta m_1^{(-1)}\hat{\gamma}_{qg}^{(0)}
                              +2\delta m_1^{(0)}\hat{\gamma}_{qg}^{(0)}
                      \Bigr]
                  + \delta m_1^{(0)}\hat{\gamma}_{qg}^{(0)}
                       \Bigl[
                               \gamma_{gg}^{(0)}
                              -\gamma_{qq}^{(0)}
                              +2\beta_0
                              +4\beta_{0,Q}
                      \Bigr]
                  -\delta m_2^{(-1)}\hat{\gamma}_{qg}^{(0)}
             \Biggr)
\N\\ &&
                 +a_{Qg}^{(3)}
                  \Biggr] 
\label{AhhhQg3} 
\\
\text{and}\nonumber\\
\hat{\hat{A}}_{gg,Q}^{(3)}&=&
                  \Bigl(\frac{\hat{m}^2}{\mu^2}\Bigr)^{3\ep/2}
                     \Biggl[
           \frac{1}{\ep^3}
             \Biggl(
                  -\frac{\gamma_{gq}^{(0)}\hat{\gamma}_{qg}^{(0)}}{6}
                                \Bigl[
                                        \gamma_{gg}^{(0)}
                                       -\gamma_{qq}^{(0)}
                                       +6\beta_0
                                       +4N_F\beta_{0,Q}
                                       +10\beta_{0,Q}
                                \Bigr]
\N\\ &&
                  -\frac{2\gamma_{gg}^{(0)}\beta_{0,Q}}{3}
                                \Bigl[
                                        2\beta_0
                                       +7\beta_{0,Q}
                                \Bigr]
                  -\frac{4\beta_{0,Q}}{3}
                                \Bigl[
                                        2\beta_0^2
                                       +7\beta_{0,Q}\beta_0
                                       +6\beta_{0,Q}^2
                                \Bigr]
             \Biggr)
\N\\ &&
          +\frac{1}{\ep^2}
             \Biggl(
                   \frac{\hat{\gamma}_{qg}^{(0)}}{6}
                                \Bigl[
                                        \gamma_{gq}^{(1)}
                                       -(2N_F-1)\hat{\gamma}_{gq}^{(1)}
                                \Bigr]
                  +\frac{\gamma_{gq}^{(0)}\hat{\gamma}_{qg}^{(1)}}{3}
                  -\frac{\hat{\gamma}_{gg}^{(1)}}{3}
                                \Bigl[
                                        4\beta_0
                                       +7\beta_{0,Q}
                                \Bigr]
\N\\ &&
                  +\frac{2\beta_{0,Q}}{3}
                                \Bigl[
                                        \gamma_{gg}^{(1)}
                                       +\beta_1
                                       +\beta_{1,Q}
                                \Bigr]
                  +\frac{2\gamma_{gg}^{(0)}\beta_{1,Q}}{3}
                           +\delta m_1^{(-1)}
                                \Bigl[
                                    -\hat{\gamma}_{qg}^{(0)}\gamma_{gq}^{(0)}
                                    -2\beta_{0,Q}\gamma_{gg}^{(0)}
                                    -10\beta_{0,Q}^2
\N\\ &&
                                    -6\beta_{0,Q}\beta_0
                               \Bigr]
             \Biggr)
          +\frac{1}{\ep}
             \Biggl(
                   \frac{\hat{\gamma}_{gg}^{(2)}}{3}
                  -2(2\beta_0+3\beta_{0,Q})a_{gg,Q}^{(2)}
                  -N_F\hat{\gamma}_{qg}^{(0)}a_{gq,Q}^{(2)}
                  +\gamma_{gq}^{(0)}a_{Qg}^{(2)}
                  +\beta_{1,Q}^{(1)} \gamma_{gg}^{(0)}
\N\\ &&
                  +\frac{\gamma_{gq}^{(0)}\hat{\gamma}_{qg}^{(0)}\zeta_2}{16}
                                \Bigl[
                                         \gamma_{gg}^{(0)}
                                       - \gamma_{qq}^{(0)}
                                       +2(2N_F+1)\beta_{0,Q}
                                       +6\beta_0
                                \Bigr]
                  +\frac{\beta_{0,Q}\zeta_2}{4}
                                \Bigl[
                                       \gamma_{gg}^{(0)}
                                      \{2\beta_0-\beta_{0,Q}\}
                                       +4\beta_0^2
\N\\ &&
                                       -2\beta_{0,Q}\beta_0
                                       -12\beta_{0,Q}^2
                                \Bigr]
                           +\delta m_1^{(-1)}
                                \Bigl[
                                   -3\delta m_1^{(-1)}\beta_{0,Q}
                                   -2\delta m_1^{(0)}\beta_{0,Q}
                                   -\hat{\gamma}_{gg}^{(1)}
                                \Bigr]
\N\\ &&
                           +\delta m_1^{(0)}
                                \Bigl[
                                   -\hat{\gamma}_{qg}^{(0)}\gamma_{gq}^{(0)}
                                   -2\gamma_{gg}^{(0)}\beta_{0,Q}
                                   -4\beta_{0,Q}\beta_0
                                   -8\beta_{0,Q}^2
                                \Bigr]
                           +2 \delta m_2^{(-1)} \beta_{0,Q}
             \Biggr)
           +a_{gg,Q}^{(3)}
                  \Biggr]~. \label{Ahhhgg3Q}
\end{eqnarray}
Here, $m$ denotes the heavy quark mass, $\varepsilon = D-4$ the dimensional parameter,
$\mu$ the factorization scale, $\zeta_l, l \in \mathbb{N}, l \geq 2$ denotes the values of the 
Riemann $\zeta$ function at integer argument, $N_F$ is the number of massless quark flavors, $\beta_i$ the 
expansion 
coefficients of the QCD $\beta$-function, $\beta_{i,Q}$ related expansion coefficients associated 
to heavy quark effects, $\gamma_{ij}^{(k)}$ the expansion coefficients of the anomalous 
dimensions, and $\delta m_k^{(l)}$ the expansion coefficients of the unrenormalized quark mass, 
cf.~\cite{{Bierenbaum:2009mv}}. The coefficients $a_{ij}^{(k)}$ denote the constant terms of 
the OMEs at $k$-loop order and $\bar{a}_{ij}^{(k)}$ the corresponding terms at $O(\varepsilon)$,
\cite{Buza:1995ie,Bierenbaum:2007qe,Bierenbaum:2008yu,Buza:1996wv,Bierenbaum:2009zt,Bierenbaum:2009mv}.

Furthermore, we use the convention
\begin{eqnarray}
\label{eq:r1}
\hat{f}(N_F)   &=&  f(N_F+1) - f(N_F)   \\
\label{eq:r2}
\tilde{f}(N_F) &=&  \frac{\displaystyle f(N_F)}{\displaystyle N_F}~.
\end{eqnarray}
The pole terms $O(1/\ep^3)$ and $O(1/\ep^2)$ contain the complete $O(a_s)$ and 
$O(a_s^2)$ 
anomalous dimensions, and the single pole term allows to obtain the terms $\propto T_F N_F$ of 
the $O(a_s^3)$ anomalous dimensions $\gamma_{qg}^{(2)}$ and $\gamma_{gg}^{(2)}$, which  in the 
case of $\gamma_{qg}^{(2)}$ is the complete anomalous dimension.

The Feynman diagrams contributing to the OMEs $A_{Qg}^{(3)}$ (\ref{AhhhQg3}) and 
$A_{gg,Q}^{(3)}$ (\ref{Ahhhgg3Q}) were generated by the code {\tt QGRAF} 
\cite{{Nogueira:1991ex}}\footnote{See Ref.~\cite{Bierenbaum:2009mv} for the implementation of 
the local operators.}. The color configurations were calculated using the package {\tt Color}  
\cite{vanRitbergen:1998pn} and the Feynman integrals were reduced to master integrals using 
the package {\tt Reduze 2} \cite{Studerus:2009ye,vonManteuffel:2012np}.\footnote{The 
package {\tt Reduze 2} uses the packages {\tt FERMAT} \cite{FERMAT} and {\tt Ginac} 
\cite{Bauer:2000cp}.}. There are different techniques available to calculate the master 
integrals, which have been summarized in Refs.~\cite{Blumlein:2017dxp} and 
\cite{Ablinger:2015tua}.
\subsection{The Anomalous Dimension \boldmath $\gamma_{qg}^{(2)}$}
\label{sec:22}

\vspace*{1mm}
\noindent
In calculating $\gamma_{qg}^{(2)}$ we have used the method of large moments 
\cite{Blumlein:2017dxp}. First the local operator insertions $(\Delta.p)^N$ are formally resummed
into propagator like terms by
\begin{eqnarray}
\label{eq:DPx}
(\Delta.p)^N \rightarrow \sum_{k=0}^\infty x^N (\Delta.p)^N = \frac{1}{1-x\Delta.p}~.
\end{eqnarray}
The system of differential equations in the variable $x$ obeyed by the master integrals 
can be obtained by integration by parts identities through Reduze 2. The master integrals 
can then be expanded into a formal Laurent-Taylor series, which can be inserted in the 
differential equations, leading to associated difference equations, with known initial conditions.
One can then exploit these difference equations to generate a large set of 
moments for each color and $\zeta$-value coefficient, expressed 
as a sequence of rational numbers. In the present approach it is required 
that the master integrals have to be known to a higher order in the dimensional parameter $\ep$.
Usually standard techniques allow to derive the corresponding expansion. In one particular case 
we had to spend some effort to obtain the initial conditions for some master integrals, which is described in 
Appendix~\ref{sec:B}.
 
With the corresponding series of rational numbers at hand, we use the algorithm of guessing \cite{GUESS} 
to derive their associated minimal recurrence. 
{Using the package {\tt Sigma}~\cite{SIG1,SIG2} this recurrence is then solved in terms 
of nested sums and the solutions are afterwards transformed to special function formats using the 
package {\tt HarmonicSums} \cite{HARMONICSUMS,Ablinger:PhDThesis,Ablinger:2011te,Ablinger:2013cf,Ablinger:2014bra}.}

The recurrences derived for the anomalous dimension are first order factorizable 
and have therefore sum- and product solutions according to difference field theory 
\cite{Karr:81,Schneider:01,Schneider:05a,
Schneider:07d,Schneider:10b,Schneider:10c,Schneider:15a,Schneider:08c}. 
Moreover, the solutions turn out to be given
even in terms of harmonic sums only. To obtain a solution conform 
with a later analytic continuation one should consider even moments only, which would 
require {a twice as large set of moments as input}. In many problems this number is so high 
that it might be difficult to reach. Alternatively, one can start by taking all moments 
and derive a first expression at general values of $N$, from which only the even moments 
are then projected. {This is algebraically possible by considering {\tt $(-1)^N$} as a variable $y$ 
and 
performing division with remainder w.r.t.\ the polynomial $1-y$ (or\ $1+y$).} In this way we obtain the $N$th 
moment of 
the single pole term. The use of also odd moments is possible in other approaches 
and recommended if a lower number of moments allows to obtain the difference equation to be 
determined. It yet corresponds to a different problem in general. {However, given a recurrence for odd and even values, one can now 
generate easily a very high number of moments using this recurrence, can extract afterwards the needed even moments and can repeat the guessing process  
with the now sufficiently large number of even moments}.

In the calculation we computed only the irreducible contributions and added the other 
terms known already as a function of $N$, cf. 
Refs.~\cite{Buza:1995ie,Bierenbaum:2007qe,Bierenbaum:2008yu,Buza:1996wv,Bierenbaum:2009zt,Bierenbaum:2009mv}.
Here the master integrals are not solved in terms of sums, but only their moments are determined and 
those are inserted into the expressions of $A_{Qg}^{(3)}$, for which, again, only moments are calculated
at first. We have then projected the irreducible contributions, using the known terms, on 
to $\gamma_{qg}^{(2)}$ using Eq.~(\ref{AhhhQg3}).

The order and degree of the recurrences associated to the different color and 
$\zeta$-factors, as well the times to guess the corresponding recurrences using {\tt Guess} \cite{GUESS}
and to solve them {using the packages {\tt Sigma} \cite{SIG1,SIG2} and {\tt HarmonicSums} 
\cite{HARMONICSUMS,Ablinger:PhDThesis,Ablinger:2011te,Ablinger:2013cf,Ablinger:2014bra}}
are summarized in Table~\ref{Tab1}. {We have generated 2000 even moments for the 
unrenormalized expression $\hat{\hat{A}}_{qg}^{(3)}$ up to its $1/\ep$ term using the package~\texttt{SolveCoupledSystem} in which the method of large moments is implemented~\cite{Blumlein:2017dxp}.}  
The computation of the moments for the master integrals took 9.30 days per kernel, where we parallelized to 15
kernels. The creation of all pole terms in $\hat{\hat{A}}_{qg}^{(3)}$ {took 10.67 days/kernel where we 
parallelized to 25 kernels.} 
The moments of the OME formed the input of the subsequent calculation. We used the even moments to 
guess the corresponding recurrences using {\tt Guess} \cite{GUESS}. The implementation of the 
guesser in {\tt Sage} \cite{SAGE,GSAGE} needs a much shorter time to find a recurrence than an earlier 
one written in {\tt Mathematica} \cite{GUESS}. The total guessing time of
$\sim 89$~sec was smaller than the corresponding one using the {\tt Mathematica}-version by a factor 75. 
The {\tt Sage}-version of the guesser does currently not report the number of 
moments used. For the most demanding example, the $C_A^2 T_F N_F$-term, 1000 moments were not 
sufficient, but 1200 moments were. In our earlier study \cite{Blumlein:2009tj}, 
slightly different recurrences were obtained, since there we determined them from even and 
odd moments. The present recurrences are somewhat shorter.

The computation time for the last two steps
to obtain $\gamma_{qg}^{(2)}$ amounted to $44.6$~min, which included the time to 
create the 
different recurrences through guessing and their solution to obtain the closed expressions as
a function of the Mellin variable $N$.

This computational time and memory requirement are much shorter than that one using the traditional 
method, where first all master integrals are calculated in explicit form and are then inserted to 
obtain the final result. The costs to determine the different master integrals as functions of $N$ would be 
much higher, if possible at all, because elliptic structures may arise at higher orders in $\ep$. 

In any case, the time needed to reduce the OME to master integrals is the longest part of the 
calculation. Using {\tt Reduze 2} 
\cite{Studerus:2009ye,vonManteuffel:2012np} the reduction time amounted to several months. Similar 
or larger run times are  expected also for other reduction programs.
\begin{table}[H]
\centering
\begin{tabular}{|r|l|r|r|r|r|}
\hline
\multicolumn{1}{|c|}{} & 
\multicolumn{1}{c|}{color/$\zeta_l$} & 
\multicolumn{1}{c|}{order} & 
\multicolumn{1}{c|}{degree} & 
\multicolumn{1}{c|}{$\tau_{\rm guess}$ [sec]} & 
\multicolumn{1}{c|}{$\tau_{\rm solve}$ [sec]}  
\\
\hline
$\gamma_{qg}^{(0)}$ & $T_F N_F$                 &  
1 &   4 &    0.16 &   0.12 \\
\hline
$\gamma_{qg}^{(1)}$ & $C_A T_F N_F$             & 
3 &  29 &    0.40 &   3.35 \\
                    & $C_F T_F N_F$             & 
3 &  26 &    0.44 &   2.73 \\
\hline
$\gamma_{qg}^{(2)}$ & $C_A^2 T_F N_F$           & 
12 & 191 &   36.41 & 1038.31 \\
                    & $C_A^2 T_F N_F \zeta_3$   &  
2 &  14 &    0.19 &   1.18 \\
                    & $C_F^2 T_F N_F$           & 
10 & 140 &   12.14 & 487.35 \\
                    & $C_F^2 T_F N_F \zeta_3$   &  
1 &   7 &    0.16 &   0.19 \\
                    & $C_F C_A T_F N_F$         & 
12 & 185 &   34.12 & 998.08 \\
                    & $C_F C_A T_F N_F \zeta_3$ &  
2 &  14 &    0.28 &   1.19 \\
                    & $C_A T_F N_F^2 $          & 
5 &  59 &    1.62 &  28.95 \\
                    & $C_F T_F N_F^2 $          & 
5 &  68 &    3.46 &  22.21 \\
\hline
\end{tabular}
\caption[]{\label{Tab1} 
\sf \small Recurrence and run-time parameters to determine the anomalous dimension 
$\gamma_{qg}^{(k)}(N)$.}
\end{table}

The $O(a_s)$ and $O(a_s^2)$ anomalous dimensions are read from the $O(1/\ep^3)$ and 
$O(1/\ep^2)$ terms and $\gamma_{qg}^{(2)}$ is determined from the $O(1/\ep)$ term.
The expressions are reduced to the polynomial basis in the harmonic sums \cite{Vermaseren:1998uu,Blumlein:1998if}  
$S_{\vec{a}}(N) \equiv S_{\vec{a}}$,
\begin{eqnarray}
S_{b,\vec{a}}(N) = \sum_{k=1}^N \frac{({\rm sign}(b))^k}{k^{|b|}} S_{\vec{a}}(k),~~
b, a_i \in \mathbb{Z} \backslash \{0\},~~N \in \mathbb{N} \backslash \{0\}, S_\emptyset = 1.
\end{eqnarray}

One obtains
\begin{eqnarray}
\gamma_{qg}^{(0)} &=& -8 \textcolor{blue}{N_F T_F} \frac{2 + N + N^2}{N (1 + N) (2 + N)}
                  \equiv -8 \textcolor{blue}{N_F T_F} p_{qg}^{(0)}(N)
\\
\gamma_{qg}^{(1)} &=&
        \textcolor{blue}{C_A N_F T_F}
        \Biggl\{
                -\frac{16 P_2}{(N-1) N^3 (N+1)^3 (N+2)^3}
                + 16 p_{qg}^{(0)}(N) \big(
                          S_1^2
                        + S_2
                        + 2 S_{-2}
                \big)
        \nonumber\\ &&
                -\frac{64 (2 N+3)}{(N+1)^2 (N+2)^2} S_1
        \Biggr\}
        \nonumber\\ &&
        +\textcolor{blue}{C_F N_F T_F}
        \Biggl\{
                -\frac{8 P_1}{N^3 (N+1)^3 (N+2)}
                +16 p_{qg}^{(0)}(N) \big(
                        - S_1^2
                        + S_2
                \big)
                +\frac{32}{N^2} S_1
        \Biggr\},
\end{eqnarray}
with the polynomials
\begin{eqnarray}
P_1 &=& 5 N^6+15 N^5+36 N^4+51 N^3+25 N^2+8 N+4
\\
P_2 &=& N^9+6 N^8+15 N^7+25 N^6+36 N^5+85 N^4+128 N^3+104 N^2+64 N+16.
\end{eqnarray}
The anomalous dimensions $\gamma_{qg}^{(0,1)}$ agree with results given in 
Refs.~\cite{Gross:1974cs,Georgi:1951sr,Floratos:1978ny,
GonzalezArroyo:1979he,Furmanski:1980cm,
Hamberg:1991qt,Ellis:1996nn,Moch:1999eb}, or with their Mellin transform.

For the three-loop anomalous dimension $\gamma_{qg}^{(2)}$, we obtain 
\begin{eqnarray}
\gamma_{qg}^{(2)} &=&
        \textcolor{blue}{C_A N_F^2 T_F^2} \Biggl\{
                -
                \frac{128(5 N^2+8 N+10)}{9 N (N+1) (N+2)}  
S_{-2}
                -\frac{64 P_8}{9 N (N+1)^2 (N+2)^2} S_1^2
        \nonumber\\ &&
                -\frac{64 P_9}{9 N (N+1)^2 (N+2)^2} S_2 
                +\frac{64 P_{25}}{27 N (N+1)^3 (N+2)^3} S_1 
                +\frac{16 P_{34}}{27 (N-1) N^4 (N+1)^4 (N+2)^4} 
        \nonumber\\ &&
                +p_{qg}^{(0)}(N) \Biggl(
                        \frac{32}{9} S_1^3
                        -\frac{32}{3} S_1 S_2
                        +\frac{64}{9} S_3
                        +\frac{128}{3} S_{-3}
                        +\frac{128}{3} S_{2,1}
                \Biggr)
        \Biggr\}
        \nonumber\\ &&
        +\textcolor{blue}{C_F N_F^2 T_F^2}
        \Biggl\{
                        \frac{32(5 N^2+3 N+2) }{3 N^2 (N+1) (N+2)}  S_2
                        +\frac{32(10 N^3+13 N^2+29 N+6) }{9 N^2 (N+1) (N+2)}  S_1^2 
        \nonumber\\ &&
                        -\frac{32 
        P_{12}}{27 N^2 (N+1)^2 (N+2)} S_1
                        +\frac{4 P_{38}}{27 (N-1) N^5 (N+1)^5 (N+2)^4} 
        \nonumber\\ &&
        +p_{qg}^{(0)}(N) \Biggl(
                                -\frac{32}{9} S_1^3
                                -\frac{32}{3} S_1 S_2
                                +\frac{320}{9} S_3
                        \Biggr)
                \Biggr\}
        \nonumber\\ &&
                +\textcolor{blue}{C_A C_F N_F T_F} \Biggl\{
                        -128 \frac{ N^3-7 N^2-6 N+4 }{N^2 (N+1)^2 (N+2)} S_{-2,1}
                        +\frac{32 P_5}{N^2 (N+1)^2 (N+2)} S_{-3}
        \nonumber\\ &&            
            +\frac{16 P_{18}}{9 (N-1) N^2 (N+1)^2 (N+2)^2} S_1^3  
                        -\frac{16 P_{24}}{9 (N-1) N^2 (N+1)^2 (N+2)^2} S_3 
        \nonumber\\ &&   
                     -\frac{8 P_{27}}{9 (N-1) N^3 (N+1)^3 (N+2)^2} S_1^2
                        +\frac{8 P_{29}}{3 (N-1) N^3 (N+1)^3 (N+2)^3} S_2
        \nonumber\\ &&
                        +\frac{P_{37}}{27 (N-1) N^5 (N+1)^5 (N+2)^4} 
                        + p_{qg}^{(0)}(N) \Biggl[
                                \left(
                                        \frac{640}{3} S_3
                                        -384 S_{2,1}
                                \right) S_1
                                +\frac{32}{3} S_1^4
        \nonumber\\ &&
                                +160 S_1^2 S_2
                                -64 S_2^2
                                +\big(
                                        192 S_1^2
                                        +64 S_2
                                \big) S_{-2}
                                +96 S_{-2}^2
                                +224 S_{-4}
                                -64 S_{2,-2}
                                +64 S_{3,1}
        \nonumber\\ &&
                                +192 S_{2,1,1}
                                -256 S_{-2,1,1}
                                -192 S_1\zeta_3
                        \Biggr]
                        -\frac{192 P_{17}}{(N-1) N^2 (N+1)^2 (N+2)^2} \zeta_3
        \nonumber\\ &&             
           +\Biggl(
                                \frac{16 P_{16}}{3 (N-1) N^2 (N+1)^2 (N+2)^2} S_2
                                +\frac{16 P_{35}}{27 (N-1) N^4 (N+1)^4 (N+2)^4}
                        \Biggr) S_1
        \nonumber\\ &&
                        +\Biggl[
                                -\frac{32 P_{15}}{N^3 (N+1)^3 (N+2)}
                                +\frac{128 \big(N^3-13 N^2-14 N-2\big)}{N^2 (N+1)^2 (N+2)}  S_1
                        \Biggr] S_{-2}
        \nonumber\\ &&
                        +\frac{96 N (N+1) p_{qg}^{(0)}(N)^2 }{N-1} S_{2,1}
        \Biggr\}
        \nonumber\\ &&
        +\textcolor{blue}{C_A^2 N_F T_F} \Biggl\{
                -\frac{64 P_{11}}{(N-1) N^2 (N+1)^2 (N+2)^2}  S_{-2,1}
                -\frac{16 P_{20}}{9 (N-1) N^2 (N+1)^2 (N+2)^2} S_3
        \nonumber\\ &&
                -\frac{32 P_{21}}{3 (N-1) N^2 (N+1)^2 (N+2)^2} S_{-3} 
                -\frac{8 P_{22}}{9 (N-1) N^2 (N+1)^2 (N+2)^2} S_1^3 
        \nonumber\\ &&
                +\frac{16 P_{32}}{9 (N-1)^2 N^3 (N+1)^3 (N+2)^3} S_1^2 
                +\frac{16 P_{33}}{9 (N-1)^2 N^3 (N+1)^3 (N+2)^3} S_2 
        \nonumber\\ &&
                -\frac{8 P_{39}}{27 (N-1)^2 N^5 (N+1)^5 (N+2)^5} 
                +p_{qg}^{(0)}(N) \Biggl[
                        -\frac{32 P_{10}}{3 (N-1) N (N+1) (N+2)} S_{2,1}
        \nonumber\\ &&                
        +\Biggl(
                                -\frac{704}{3} S_3
                                +128 S_{2,1}
                                +512 S_{-2,1}
                        \Biggr) S_1
                        -512 S_{-3} S_1
                        -\frac{16}{3} S_1^4
                        -160 S_1^2 S_2
                        -16 S_2^2
                        -32 S_4
        \nonumber\\ &&
                        +\Biggl(
                                -192 S_1^2
                                +320 S_2
                        \Biggr) S_{-2}
                        -96 S_{-2}^2
                        +96 S_{-4}
                        -448 S_{2,-2}
                        -128 S_{3,1}
                        +512 S_{-3,1}
        \nonumber\\ &&
                        -768 S_{-2,1,1}
                        +192 S_1 \zeta_3
                \Biggr]
                +\frac{96 (N-2) (N+3) P_4}{(N-1) N^2 (N+1)^2 (N+2)^2}  \zeta_3
        \nonumber\\ &&
                +\Biggl(
                        \frac{8 P_{19}}{3 (N-1) N^2 (N+1)^2 (N+2)^2} S_2
                        -\frac{8 P_{36}}{27 (N-1)^2 N^4 (N+1)^4 (N+2)^4} 
                \Biggr) S_1
        \nonumber\\ &&
                +\Biggl(
                        -\frac{64 P_{13}}{(N-1) N^2 (N+1)^2 (N+2)^2} S_1
                        +\frac{32 P_{30}}{9 (N-1) N^3 (N+1)^3 (N+2)^3} 
                \Biggr) S_{-2}
        \Biggr\}
        \nonumber\\ &&
        + \textcolor{blue}{C_F^2 N_F T_F} \Biggl\{
                \frac{P_{31}}{N^5 (N+1)^5 (N+2)}
                -\frac{8 P_3}{3 N^2 (N+1)^2 (N+2)} S_1^3
                -\frac{16 P_6}{3 N^2 (N+1)^2 (N+2)} S_3 
        \nonumber\\ && 
                +\frac{64 P_{14}}{N^3 (N+1)^2 (N+2)} S_{-2} 
                -\frac{8 P_{23}}{N^3 (N+1)^3 (N+2)} S_1^2
                +\frac{8 P_{26}}{N^3 (N+1)^3 (N+2)} S_2
        \nonumber\\ &&
                + p_{qg}^{(0)}(N) \Biggl[
                        \Biggl(
                                -\frac{704}{3} S_3
                                +256 S_{2,1}
                        \Biggr) S_1
                        -256 S_{-3} S_1
                        -\frac{16}{3} S_1^4
                        -48 S_2^2
                        -160 S_4
                        -64 S_{-2}^2
        \nonumber\\ &&
                        -192 S_{-4}
                        -\frac{128}{N (N+1)} S_{2,1}
                        -128 S_{2,-2}
                        +64 S_{3,1}
                        +256 S_{-3,1}
                        -192 S_{2,1,1}
                \Biggr]
        \nonumber\\ &&
                +\frac{96 (N-1) \big(
                        3 N^2+3 N-2\big) }{N^2 (N+1)^2} \zeta_3
                - 256 \frac{2-N+N^2}{N^2 (N+1) (N+2)}
                        \left[S_{-2} S_1
                        - S_{-2,1}\right]
        \nonumber\\ &&
                +\Biggl(
                        -\frac{8 P_{28}
                        }{N^4 (N+1)^4 (N+2)}
                        -\frac{8 P_7 }{N^2 (N+1)^2 (N+2)}  S_2
                \Biggr) S_1
                -\frac{128 (N-1)}{(N+1)^2 (N+2)} S_{-3}
        \Biggr\},
\label{eq:gqg2}
\end{eqnarray}
where
\begin{eqnarray}
P_3 &=& 3 N^4-6 N^3-37 N^2-52 N-20
\\
P_4 &=& 3 N^4+6 N^3+7 N^2+4 N+4
\\
P_5 &=& 3 N^4+8 N^3-5 N^2-6 N+8
\\
P_6 &=& 3 N^4+30 N^3-13 N^2+8 N+4
\\
P_7 &=& 3 N^4+34 N^3+67 N^2+60 N+12
\\
P_8 &=& 5 N^4+20 N^3+41 N^2+49 N+20
\\
P_9 &=& 5 N^4+26 N^3+47 N^2+43 N+20
\\
P_{10} &=& 11 N^4+22 N^3+13 N^2+2 N+24
\\
P_{11} &=& 33 N^4+54 N^3+9 N^2-4 N+4
\\
P_{12} &=& 47 N^4+145 N^3+426 N^2+412 N+120
\\
P_{13} &=& 2 N^5-23 N^4-32 N^3+13 N^2+4 N-12
\\
P_{14} &=& 2 N^5+4 N^4-N^3+N^2-2 N+8
\\
P_{15} &=& 3 N^5-40 N^4-87 N^3-54 N^2-10 N+12
\\
P_{16} &=& N^6+21 N^5-72 N^4-335 N^3-259 N^2+128 N+84
\\
P_{17} &=& 3 N^6+9 N^5-5 N^4-25 N^3-14 N^2-16
\\
P_{18} &=& 10 N^6+12 N^5-117 N^4-122 N^3+137 N^2+140 N+84
\\
P_{19} &=& 11 N^6-15 N^5+345 N^4+371 N^3-164 N^2+172 N+144
\\
P_{20} &=& 11 N^6+33 N^5-456 N^4-751 N^3-1001 N^2-872 N-996
\\
P_{21} &=& 11 N^6+33 N^5-114 N^4-247 N^3-263 N^2-176 N-108
\\
P_{22} &=& 11 N^6+33 N^5-87 N^4-13 N^3+268 N^2+28 N+48
\\
P_{23} &=& 14 N^6+43 N^5+77 N^4+149 N^3+171 N^2+122 N+40
\\
P_{24} &=& 19 N^6+21 N^5+477 N^4+967 N^3+272 N^2-76 N+624
\\
P_{25} &=& 19 N^6+124 N^5+492 N^4+1153 N^3+1362 N^2+712 N+152
\\
P_{26} &=& 26 N^6+93 N^5+179 N^4+227 N^3+109 N^2+54 N+24
\\
P_{27} &=& 8 N^8-7 N^7+408 N^6+320 N^5+840 N^4+2807 N^3+1804 N^2+1596 N+576
\\
P_{28} &=& 17 N^8+43 N^7-33 N^6-297 N^5-502 N^4-442 N^3-354 N^2-216 N-56
\\
P_{29} &=& 36 N^9+125 N^8+142 N^7+340 N^6+1040 N^5+2095 N^4+2930 N^3+2900 N^2
\nonumber\\ &&
+1816 N+384
\\
P_{30} &=& 94 N^9+597 N^8+1508 N^7+2086 N^6+1517 N^5+1381 N^4+2731 N^3+3802 N^2
\nonumber\\ &&
+2916 N+648
\\
P_{31} &=& -5 N^{10}-25 N^9-64 N^8-434 N^7-913 N^6-1609 N^5-3186 N^4-3276 N^3
\nonumber\\ &&
-1784 N^2-608 N-128
\\
P_{32} &=& 67 N^{10}+317 N^9+656 N^8+839 N^7+100 N^6-250 N^5+1089 N^4+1326 N^3-40 N^2
\nonumber\\ &&
-1080 N-432
\\
P_{33} &=& 85 N^{10}+491 N^9+968 N^8+713 N^7-536 N^6-130 N^5+1467 N^4+1158 N^3+176 N^2
\nonumber\\ &&
-1368 N-432
\\
P_{34} &=& 165 N^{12}+1485 N^{11}+5194 N^{10}+8534 N^9+3557 N^8-8899 N^7-10364 N^6+6800 N^5
\nonumber\\ &&
+25896 N^4+30864 N^3+19904 N^2+7296 
N+1152
\\
P_{35} &=& 476 N^{12}+4677 N^{11}+19859 N^{10}+49847 N^9+86847 N^8+107958 N^7+87334 N^6
\nonumber\\ &&
+55746 N^5+86344 N^4+144544 N^3+128352 
N^2+56160 N+10368
\\
P_{36} &=& 475 N^{13}+3683 N^{12}+15244 N^{11}+36480 N^{10}+28631 N^9-53577 N^8
\nonumber\\ &&
-150844 N^7-150848 N^6-67642 N^5-35538 
N^4-43808 N^3-14256 N^2-864 N
\nonumber\\ &&
-2592
\\
P_{37} &=& -1251 N^{14}-12510 N^{13}-95893 N^{12}-532884 N^{11}-1824073 N^{10}-3974206 N^9
\nonumber\\ &&
-6099363 
N^8-7385376 N^7-7719308 N^6-7064880 N^5-5220560 N^4-2853632 N^3
\nonumber\\ &&
-1178688 N^2-357120 N-55296
\\
P_{38} &=& 99 N^{14}+990 N^{13}+4925 N^{12}+17916 N^{11}+46649 N^{10}+72446 N^9+32283 N^8
\nonumber\\ &&
-95592 N^7-267524 N^6-479472 N^5
-586928 N^4-455168 N^3-269760 N^2
\nonumber\\ &&
-122112 N-27648
\\
P_{39} &=& 741 N^{16}+8151 N^{15}+35407 N^{14}+72811 N^{13}+44435 N^{12}-94609 N^{11}-132463 N^{10}
\nonumber\\ &&
+214525 N^9+800336 N^8
+1203482 N^7+1303768 N^6+1266904 N^5+1100416 N^4
\nonumber\\ &&
+773216 N^3+431232 N^2+160128 N+27648.
\end{eqnarray}
The anomalous dimension $\gamma_{qg}^{(2)}$, Eq.~(\ref{eq:gqg2}), agrees with the 
corresponding expression given in Ref.~\cite{Vogt:2004mw}. The terms $\propto N_F^2$ were confirmed
before in \cite{Ablinger:2010ty}.
\subsection{The Anomalous Dimension \boldmath $\gamma_{gg}^{(2),\rm N_F}$}
\label{sec:23}

\vspace*{1mm}
\noindent
The anomalous dimensions $\gamma_{gg}^{(0,1)}(N)$ can be obtained from the pole terms $O(1/\ep^3)$ 
and
$O(1/\ep^2)$ of massive OME $A_{gg,Q}^{(3)}$, while the single pole term $O(1/\ep)$ contains  
$\gamma_{gg}^{(2),\rm N_F}(N)$, by virtue of being $\propto T_F$. We compute the contributions to 
$A_{gg,Q}^{(3)}$ up to the $1/\ep$ terms in the traditional way, i.e. the contributing master 
integrals are first calculated in explicit form. A subset of master integrals can be obtained
using representations through (generalized) hypergeometric functions \cite{HYP,SLATER} or using 
Mellin-Barnes integrals \cite{MB1,MB}. {The resulting definite sums have been simplified afterwards with symbolic summation
using the packages {\tt Sigma} \cite{SIG1,SIG2}, {\tt EvaluateMultiSums} \cite{EMSSP}, 
and {\tt HarmonicSums} 
\cite{HARMONICSUMS,Ablinger:PhDThesis,Ablinger:2011te,Ablinger:2013cf,Ablinger:2014bra}.
A further large set is calculated using the method of 
differential equations \cite{DEQ,Ablinger:2015tua}. Here we use the algorithm described in 
Ref.~\cite{Ablinger:2015tua}, which is available in \texttt{SolveCoupledSystem} using the packages 
\texttt{Sigma}, \texttt{HarmonicSums}, \texttt{EvaluateMultiSums}, {\tt SumProduction} 
\cite{EMSSP} and~\texttt{OreSys}~\cite{Gerhold:2002}. This algorithm 
allows to solve the differential equations analytically in the 
one-parameter case without the need to use any specific basis of master integrals. In the present 
calculation of anomalous dimensions, the corresponding difference equations are factorizing in 
first order and can be solved analytically using the package \texttt{Sigma}.
In a single case the computation time turned out to be large. Therefore, we used the 
Almkvist-Zeilberger theorem~\cite{AZ} and our packages~\texttt{MultiIntegrate}~\cite{Ablinger:PhDThesis} to 
derive a linear recurrence. Solving the recurrence with \texttt{Sigma} and \texttt{HarmonicSums} provided then an immediate solution.}
The master integrals are obtained as expressions in the variable $x$, see Eq.~(\ref{eq:DPx}).
The solution in Mellin $N$-space is then given by  the $N$th expansion coefficient of these 
expressions, which is found by a routine of {\tt HarmonicSums}.

We obtain the following expressions for $\gamma^{(0,1),N_F}_{gg}$:
\begin{eqnarray}
\gamma^{(0)}_{gg} &=&
        \textcolor{blue}{C_A} \left[
                -\frac{2\left(11 N^4+22 N^3+13 N^2+2 N+24\right)}{3(N-1) N (N+1) (N+2)} 
                +8 S_1
        \right]
        + \textcolor{blue}{T_F N_F} \frac{8}{3} 
\\
\gamma^{(1)}_{gg} &=&
        \textcolor{blue}{C_A^2} \Biggl\{
                \frac{64 \big(N^2+N+1\big) }{(N-1) N (N+1) (N+2)} S_2
                -\frac{4 P_{43}}{9(N-1)^2 N^3 (N+1)^3 (N+2)^3} 
        \nonumber\\ &&
                +\Biggl[
                        \frac{8 P_{42}}{9(N-1)^2 N^2 (N+1)^2 (N+2)^2} 
                        -32 S_2
                \Biggr] S_1
                -16 S_3
        \nonumber\\ && 
                +\Biggl[\frac{64 \big(N^2+N+1\big)}{(N-1) N (N+1) (N+2)} 
                        -32 S_1
                \Biggr] S_{-2}
                -16 S_{-3}
                +32 S_{-2,1}
        \Biggr\}
        \nonumber\\ &&
        +\textcolor{blue}{C_A N_F T_F} \Biggl\{
                \frac{32 P_{40}}{9(N-1) N^2 (N+1)^2 (N+2)} 
                -\frac{160}{9} S_1
        \Biggr\}
        \nonumber \\ &&
        +\textcolor{blue}{C_F N_F T_F} \frac{8 P_{41}}{(N-1) N^3 (N+1)^3 (N+2)}, 
\end{eqnarray}
with
\begin{eqnarray}
P_{40} &=& 3 N^6+9 N^5+22 N^4+29 N^3+41 N^2+28 N+6
\\
P_{41}&=&N^8+4 N^7+8 N^6+6 N^5-3 N^4-22 N^3-10 N^2-8 N-8
\\
P_{42}&=&67 N^8+268 N^7+134 N^6-392 N^5-109 N^4+844 N^3+772 N^2-144 N-144
\\
P_{43}&=&48 N^{11}+336 N^{10}+1225 N^9+3030 N^8+4744 N^7+4514 N^6+1663 N^5-1384 N^4
\nonumber\\ &&
-1248 N^3+560 N^2+1488 N+576.
\end{eqnarray}
The anomalous dimensions $\gamma_{gg}^{(0,1)}$
agree with results given in \cite{Gross:1974cs,Georgi:1951sr,Floratos:1978ny,
GonzalezArroyo:1979he,Furmanski:1980cm,Hamberg:1991qt,Ellis:1996nn,Moch:1999eb}, or
with their Mellin transform.

For the contribution $\propto N_F$ to $\gamma^{(2)}_{gg}$ at three-loop order, $\gamma^{(2),N_F}_{gg}$, we 
obtain:
\begin{eqnarray}
\gamma^{(2),N_F}_{gg} &=& 
        \textcolor{blue}{C_A N_F^2 T_F^2} 
        \Biggl\{
                -\frac{16 P_{52}}{27(N-1) N^2 (N+1)^2 (N+2)} S_1
                -\frac{4 P_{59}}{27(N-1) N^3 (N+1)^3 (N+2)} 
        \Biggr\}
        \nonumber\\ &&
        +\textcolor{blue}{C_F N_F^2 T_F^2} 
        \Biggl\{
         \frac{32\big(N^2+N+2\big)^2}{3(N-1) N^2 (N+1)^2 (N+2)} S_1^2
        -\frac{32\big(N^2+N+2\big)^2}{(N-1) N^2 (N+1)^2 (N+2)} S_2
        \nonumber\\ &&
        +\frac{64 P_{51}}{9(N-1) N^3 (N+1)^3 (N+2)} S_1
        -\frac{8 P_{61}}{27(N-1) N^4 (N+1)^4 (N+2)} 
        \Biggr\}
        \nonumber\\ &&
        +\textcolor{blue}{C_F C_A N_F T_F} 
        \nonumber\\ && \times
        \Biggl\{
        -\frac{32\big(N^2+N-34\big)\big(N^2+N+2\big)}{3(N-1) N^2 (N+1)^2 (N+2)} S_3 
        +\frac{32\big(N^2+N+2\big)^2}{3(N-1) N^2 (N+1)^2 (N+2)} S_1^3
        \nonumber\\ &&
        +\frac{64  \big(N^2+N+2\big)^2}{(N-1) N^2 (N+1)^2 (N+2)} S_{2,1}
        +\frac{192 \big(N^2+N+2\big)\big(N^2+N+4\big)}{(N-1) N^2 (N+1)^2 (N+2)} S_{-3}
        \nonumber\\ &&
        -\frac{128 \big(N^2+N+2\big)\big(N^2+N+8\big)}{(N-1) N^2 (N+1)^2 (N+2)} S_{-2,1}
        -\frac{8 \big(N^2+N+2\big) P_{53}}{3(N-1)^2 N^3 (N+1)^3 (N+2)^2} S_1^2
        \nonumber\\ &&
        -\frac{8 P_{57}}{(N-1)^2 N^3 (N+1)^3 (N+2)^2} S_2
        +\frac{2 P_{66}}{27 (N-1)^2 N^5 (N+1)^5 (N+2)^4} 
        \nonumber\\ &&
        -\frac{64 \big(N^2+N+4\big)\big(N^2+N+6\big)}{(N-1) N^2 (N+1)^2 (N+2)} \zeta_3
        +\Biggl[
        -\frac{32 \big(N^2+N+2\big)^2}{(N-1) N^2 (N+1)^2 (N+2)} S_2
        \nonumber\\ &&
        -\frac{8 P_{62}}{9(N-1)^2 N^4 (N+1)^4 (N+2)^3} +128 \zeta_3
        \Biggr] S_1
        +\Biggl[
         \frac{128 P_{44}}{(N-1) N^2 (N+1)^2 (N+2)} S_1
        \nonumber\\ &&
        -\frac{64 P_{58}}{(N-1)^2 N^3 (N+1)^3 (N+2)^2}
                \Biggr] S_{-2}
        \Biggr\}
        \nonumber\\ &&
        +\textcolor{blue}{C_A^2 N_F T_F} \Biggl\{
                \frac{32 P_{50}}{9(N-1) N^2 (N+1)^2 (N+2)}  S_2
                +\frac{32 P_{54}}{9 (N-1) N^2 (N+1)^2 (N+2)} S_{-3}
        \nonumber\\ &&
                -\frac{64 P_{54}}{9(N-1) N^2 (N+1)^2 (N+2)} S_{-2,1}
                +\frac{2 P_{65}}{27 (N-1)^2 N^4 (N+1)^5 (N+2)^4} 
        \nonumber\\ &&
                +\frac{128 P_{45}}{(N-1) N^2 (N+1)^2 (N+2)} \zeta_3
                +\Biggl[
                        -\frac{8 P_{63}}{27(N-1)^2 N^4 (N+1)^4 (N+2)^3} 
        \nonumber\\ &&
                        +\frac{1280}{9} S_2
                        -\frac{64}{3} S_3
                        -128 \zeta_3
                \Biggr] S_1
                +\frac{16 P_{48}}{9 N^2 (N+1)^2} S_3
                +\Biggl[
                         \frac{64 P_{55}}{9(N-1) N^2 (N+1)^2 (N+2)} S_1
        \nonumber\\ &&
                        -\frac{32 P_{60}}{9(N-1)^2 N^3 (N+1)^3 (N+2)^2} 
                \Biggr] S_{-2}
                +\frac{64}{3} S_{-2}^2
        \Biggr\}
        \nonumber\\ &&
        +\textcolor{blue}{C_F^2 N_F T_F}
        \Biggl\{
                -\frac{256 \big(N^2+N+2\big)}{(N-1) N^2 (N+1)^2 (N+2)}  S_{-3}
                +\frac{512 \big(N^2+N+2\big)}{(N-1) N^2 (N+1)^2 (N+2)}  S_{-2,1}
        \nonumber\\ &&
                -\frac{32\big(N^2+N+2\big)^2}{3(N-1) N^2 (N+1)^2 (N+2)} S_1^3
                -\frac{64 \big(N^2+N+2\big)^2}{(N-1) N^2 (N+1)^2 (N+2)} S_{2,1}
        \nonumber\\ &&
                +\frac{32\big(N^2+N+2\big)\big(7 N^2+7 N-10\big)}{3(N-1) N^2 (N+1)^2 (N+2)} S_3
                +\frac{8 \big(N^2+N+2\big) P_{46}}{(N-1) N^3 (N+1)^3 (N+2)} S_1^2
        \nonumber\\ &&
                -\frac{8 \big(N^2+N+2\big) P_{47}}{(N-1) N^3 (N+1)^3 (N+2)} S_2
                -\frac{4 P_{64}}{3(N-1) N^5 (N+1)^5 (N+2)^4} 
        \nonumber\\ &&
               +\Biggl[
                        \frac{32 \big(N^2+N+2\big)^2}{(N-1) N^2 (N+1)^2 (N+2)} S_2
                        -\frac{32 P_{56}}{(N-1) N^4 (N+1)^4 (N+2)} 
                \Biggr] S_1
        \nonumber\\ &&
               +\Biggl[-\frac{512 \big(N^2+N+2\big)}{(N-1) N^2 (N+1)^2 (N+2)} S_1
                       -\frac{64 P_{49}}{(N-1) N^3 (N+1)^3 (N+2)} 
                \Biggr] S_{-2}
        \nonumber\\ &&
               -\frac{192 \big(N^2+N+2\big)}
                {N^2 (N+1)^2} \zeta_3
        \Biggr\},
\label{eq:ggg2}
\end{eqnarray}
where the polynomials are
\begin{eqnarray}
P_{44}&=&N^4+2 N^3+7 N^2+6 N+16
\\
P_{45}&=&2 N^4+4 N^3+7 N^2+5 N+6
\\
P_{46}&=&5 N^4+10 N^3+25 N^2+20 N+4
\\
P_{47}&=&15 N^4+14 N^3+35 N^2+20 N-4
\\
P_{48}&=&40 N^4+80 N^3+73 N^2+33 N+54
\\
P_{49}&=&N^6+3 N^5+N^4-3 N^3-26 N^2-24 N-16
\\
P_{50}&=&3 N^6+9 N^5-113 N^4-241 N^3-274 N^2-152 N-24
\\
P_{51}&=&4 N^6+3 N^5-50 N^4-129 N^3-100 N^2-56 N-24
\\
P_{52}&=&8 N^6+24 N^5-19 N^4-78 N^3-253 N^2-210 N-96
\\
P_{53}&=&17 N^6+51 N^5+99 N^4+113 N^3-32 N^2-80 N-24
\\
P_{54}&=&20 N^6+60 N^5+11 N^4-78 N^3-121 N^2-72 N-108
\\
P_{55}&=&20 N^6+60 N^5+11 N^4-78 N^3-85 N^2-36 N-108
\\
P_{56}&=&N^8+11 N^7+35 N^6+73 N^5+114 N^4+118 N^3+44 N^2-12 N-8
\\
P_{57}&=&3 N^8-108 N^6-358 N^5-375 N^4-74 N^3+40 N^2+184 N+112
\\
P_{58}&=&3 N^8+10 N^7+13 N^6+N^5+28 N^4+81 N^3+4 N^2-12 N-32
\\
P_{59}&=&87 N^8+348 N^7+848 N^6+1326 N^5+2609 N^4+3414 N^3+2632 N^2
\nonumber\\ &&
+1088 N+192
\\
P_{60}&=&131 N^8+524 N^7+691 N^6+239 N^5-848 N^4-1483 N^3-586 N^2+108 N+360
\\
P_{61}&=&33 N^{10}+165 N^9+256 N^8-542 N^7-3287 N^6-8783 N^5-11074 N^4-9624 N^3
\nonumber\\ &&
-5960 N^2-2112 N-288
\\
P_{62}&=&165 N^{13}+1320 N^{12}+3643 N^{11}+2530 N^{10}-7013 N^9-17220 N^8-21563 N^7
\nonumber\\ &&
-28630 N^6-24624 N^5+448 N^4+16320 N^3+18368 N^2+11328 N+3456
\\
P_{63}&=&418 N^{13}+3344 N^{12}+10127 N^{11}+15571 N^{10}+19386 N^9+43782 N^8+101537 N^7
\nonumber\\ &&
+138421 N^6+87796 N^5-1750 N^4-38040 N^3-26352 N^2-9504 N-2592
\\
P_{64}&=&3 N^{15}+36 N^{14}+57 N^{13}-906 N^{12}-7029 N^{11}-27744 N^{10}-72309 N^9-131682 N^8
\nonumber\\ &&
-169098 N^7-146808 N^6-69960 N^5+4608 N^4+30240 N^3+21120 N^2+8064 N
\nonumber\\ &&
+1536
\\
P_{65}&=&699 N^{15}+7689 N^{14}+48311 N^{13}+214817 N^{12}+656205 N^{11}+1354267 N^{10}
\nonumber\\ &&
+1891493 N^9+1781363 N^8+1206028 N^7+908824 N^6+1133136 N^5+1376048 N^4
\nonumber\\ &&
+1169728 N^3+633088 N^2+198144 N+27648
\\
P_{66}&=&723 N^{16}+7953 N^{15}+41771 N^{14}+132305 N^{13}+231993 
N^{12}+32347 N^{11}-890103 N^{10}
\nonumber\\ &&
-2142125 N^9-2309864 N^8-970928 N^7+346920 N^6+253856 N^5
\nonumber\\ &&
-691200 N^4-1126528 N^3-946560 N^2-486144 N-110592.
\end{eqnarray}
The anomalous dimension $\gamma_{gg}^{(2), \rm N_F}$, Eq.~(\ref{eq:ggg2}), agrees with the 
corresponding expression given in Ref.~\cite{Vogt:2004mw}. We have confirmed the leading $N_F$-terms
$\propto N_F^2$ in \cite{Blumlein:2012vq,Ablinger:2014uka}, which were also given in \cite{Bennett:1997ch}.
\section{Conclusions}
\label{sec:3}

\vspace*{1mm}
\noindent
We have computed the 3-loop anomalous dimension $\gamma_{qg}^{(2)}(N)$, the 
contributions $\propto N_F$ for $\gamma_{gg}^{(2)}(N)$ and the associated splitting 
functions in a massive calculation, which is fully independent of the earlier 
computation in Ref.~\cite{Vogt:2004mw}. We agree with the previous results. 
In the case of $\gamma_{qg}^{(2)}(N)$, we have performed a fully automatic calculation
generating the Feynman diagrams, reducing them to master integrals and calculating 
them using differential and difference equations. We used formal Taylor series representations 
in terms of the subsidiary parameter $x$ resumming the local operator insertions.
The new method of high Mellin moments \cite{Blumlein:2017dxp} has been used. Here the moments 
are calculated recursively using the difference equation systems associated to the differential 
equations. We did not compute the master integrals as explicit functions of 
$N$, but kept them as vectors of moments. The method of guessing has then been used 
to obtain one difference 
equation for each color/$\zeta$-value factor. All difference equations obtained 
factorize at first order and one finally obtains a representation in harmonic sums. 

The contributions $\propto N_F$ to $\gamma_{gg}^{(2)}(N)$ have been calculated 
using  more traditional methods. After the reduction to the master integrals, we have 
calculated these as functions of the Mellin variable $N$. The corresponding 
techniques are discussed in detail in Ref.~\cite{Ablinger:2015tua} and include
(generalized) hypergeometric and associated higher transcendental function 
representations \cite{HYP,SLATER}, the method of hyperlogarithms 
\cite{Brown:2008um,Ablinger:2014yaa,Panzer:2014caa}, differential equation techniques 
\cite{DEQ,Ablinger:2015tua}, and integration using the Almkvist-Zeilberger theorem 
\cite{AZ,Ablinger:PhDThesis}. 
The final solution has then been obtained by inserting 
the master integrals and extracting $\gamma_{gg}^{(2), \rm N_F}(N)$ from the single 
pole term of $\hat{\hat{A}}_{gg}^{(3)}$. In this process, quite a series 
of sums cancel, including all non-harmonic sums. 

The following {\tt Mathematica}, {\tt Fortran}, and {\tt FORM} \cite{Vermaseren:2000nd} packages,
see also Ref.~\cite{Bierenbaum:2009mv}\footnote{Here the renormalization of the massive operator 
matrix elements has been carried out in a {\tt maple} code.},
were instrumental in performing the present calculation: 
{\tt QGRAF} \cite{Nogueira:1991ex,Bierenbaum:2009mv}, 
{\tt Color} \cite{vanRitbergen:1998pn}, {\tt MB} \cite{MB},
{\tt Reduze 2} \cite{Studerus:2009ye,vonManteuffel:2012np},
{\tt FERMAT} \cite{FERMAT}, 
{\tt Ginac} \cite{Bauer:2000cp},
{\tt Sigma} \cite{SIG1,SIG2},
{\tt EvaluateMultiSums} and {\tt SumProduction} \cite{EMSSP},
{{\tt SolveCoupledSystems} \cite{DEQ,Ablinger:2015tua,Blumlein:2017dxp}, \texttt{OreSys}~\cite{Gerhold:2002}},
{\tt HarmonicSums} 
\cite{HARMONICSUMS,Ablinger:PhDThesis,Ablinger:2011te,Ablinger:2013cf,Ablinger:2014bra}
and {\tt Guess} \cite{GUESS,GSAGE}.

The universality of the QCD anomalous 
dimension allows to compute them within various setups. 
In the present calculation, they were obtained from the pole structure of massive OMEs. The calculation of these
OMEs is part of an ongoing project with the final goal to compute the massive Wilson coefficients for
deep-inelastic scattering in the region $Q^2 \geq m^2$.

The anomalous dimensions and splitting functions presented in this paper, together with our earlier 
results in Refs.~\cite{Ablinger:2014lka,Ablinger:2014vwa,Ablinger:2010ty,Blumlein:2012vq,Ablinger:2014uka,
Ablinger:2014nga}, are given in {\tt Mathematica.m} 
format in the attachment, together with the expressions for $I^{\rm B5a}_{\rm 101}$ for 
$D=4+\ep$ to $D=30+\ep$ up to $O(\ep^6)$ presented in appendix~\ref{sec:B}. Their conversion 
to  {\tt maple} or  {\tt Form}-inputs \cite{Vermaseren:2000nd} is straightforward.

\appendix
\section{The Splitting Functions}
\label{sec:A}

\vspace*{1mm}
\noindent
In the following we present the splitting functions $P_{qg}^{(i)}$ and $P_{qg}^{(i)}, 
i = 0,1,2$ in $x$-space. They are obtained as a Mellin-inversion from the even moments of 
$\gamma_{qg}^{(i)}$  and $\gamma_{gg}^{(i)}$ via
\begin{eqnarray}
\gamma_{kl}(N) = - \int_0^1 dx x^{N-1} P_{kl}(x)~.
\end{eqnarray}
The splitting functions to 3-loop order can be expressed in terms of  harmonic polylogarithms 
$H_{\vec{a}}(x) \equiv H_{\vec{a}}$ over the alphabet
$f_b(x) \in \{1/x,1/(1-x),1/(1+x)\}$ \cite{Remiddi:1999ew}. They are defined by
\begin{eqnarray}
H_{b,\vec{a}}(x) = \int_0^x dy f_b(y) H_{\vec{a}}(y),~~~b,a_i \in \{0,1,-1\},~~~H_\emptyset = 1.
\end{eqnarray}

The splitting functions are~:
\begin{eqnarray}
P_{qg}^{(0)} &=& 
        8 \textcolor{blue}{N_F T_F} \big( x^2 + (1-x)^2 \big)
        \equiv 8 \textcolor{blue}{N_F T_F} p_{qg}^{(0)}(x),
\\
P_{qg}^{(1)} &=&
        -\textcolor{blue}{C_A N_F T_F} \Biggl\{
                \frac{16}{9} \frac{
                        218 x^3-225 x^2+18 x-20}{x}
                + 32 p_{qg}^{(0)}(-x) \big(
                          H_{-1} H_0
                        - H_{0,-1}
                \big)
        \nonumber\\ &&
                -\frac{16}{3} \big(
                        44 x^2+24 x+3\big) H_0
                +16 (2 x+1) H_0^2
                +64 (x-1) x H_1
                +16 p_{qg}^{(0)}(x) H_1^2
                +64 x \zeta_2
        \Biggr\}
        \nonumber\\ &&
        -\textcolor{blue}{C_F N_F T_F} \Biggl\{
                -8 \big(
                        20 x^2-29 x+14\big)
                + p_{qg}^{(0)}(x) \big(
                        -32 H_0 H_1
                        -16 H_1^2
                        +32 \zeta_2
                \big)
        \nonumber\\ &&
                -8 \big(
                        8 x^2-4 x+3\big) H_0
                -8 \big(
                        4 x^2-2 x+1\big) H_0^2
                -64 (x-1) x H_1
        \Biggr\},
\end{eqnarray}
\begin{eqnarray}
P_{qg}^{(2)} &=& 
        - \Biggl\{
                \textcolor{blue}{C_F C_A N_F T_F} 
        \Biggl\{
                         \frac{16 \big(52 x^3-53 x^2+7 x+4\big)}{9 x}  H_1^3
        \nonumber\\ &&
                        +\frac{15988 x^3-13818 x^2-1429 x-880}{3 x} 
                        + p_{qg}^{(0)}(x) \Biggl[
                                 \big(
                                        16 H_0^3
                                        -576 H_0 \zeta_2
                                 \big) H_1
        \nonumber\\ &&
                                +\big(
                                        -16 H_0^2
                                        -192 \zeta_2
                                \big) H_1^2
                                -32 H_0 H_1^3
                                +\frac{32}{3} H_1^4
                                -192 H_0 H_1 H_{0,-1}
                                +\big(
                                        64 H_0 H_1
        \nonumber\\ &&
                                        +192 H_1^2
                                        -192 H_{0,-1}
                                \big) H_{0,1}
                                +384 H_1 H_{0,0,-1}
                                +320 H_1 H_{0,0,1}
                                -192 H_1 H_{0,1,1}
        \nonumber\\ &&
                                +320 H_{0,-1,0,1}
                        \Biggr]
                        + p_{qg}^{(0)}(-x) \Biggl[
                                \big(
                                        64 H_{-1}^3
                                        -576 H_{-1} \zeta_2
                                \big) H_0
                                -96 H_{-1}^2 H_0^2
                                +\frac{176}{3} H_{-1} H_0^3
        \nonumber\\ &&
                                +\big(
                                        -192 H_{-1}^2
                                        +192 H_{-1} H_0
                                \big) H_{0,-1}
                                +\big(
                                        -192 H_{-1}^2
                                        +128 H_{-1} H_0
                                \big) H_{0,1}
        \nonumber\\ &&
                                +384 H_{-1} H_{0,-1,-1}
                                +384 H_{-1} H_{0,-1,1}
                                +128 H_{-1} H_{0,0,1}
                                +384 H_{-1} H_{0,1,-1}
        \nonumber\\ &&
                                +128 H_{-1} H_{0,1,1}
                                -384 H_{0,-1,-1,-1}
                                -384 H_{0,-1,-1,1}
                                -384 H_{0,-1,1,-1}
                                -128 H_{0,-1,1,1}
        \nonumber\\ &&
                                -128 H_{0,0,-1,1}
                                -128 H_{0,0,1,-1}
                                -384 H_{0,1,-1,-1}
                                -128 H_{0,1,-1,1}
                                -128 H_{0,1,1,-1}
        \nonumber\\ &&
                                +288 H_{-1}^2 \zeta_2
                        \Biggr]
                        +\Biggl[
                                -
                                \frac{16 \big(
                                        560 x^3-448 x^2-25 x-48\big)}{3 x}
                                -544 p_{qg}^{(0)}(-x) H_{-1}
        \nonumber\\ &&
                                -32 \big(
                                        4 x^2-82 x-11\big) H_0
                                -1248 p_{qg}^{(0)}(x) H_1
                        \Biggr] \zeta_3
                        +\Biggl[
                                -\frac{4\big(
                                        11332 x^2+24569 x+2696\big)}{27}
        \nonumber\\ && 
                                +32 (x+1) (18 x-37) H_{-1}
                                +32 (x+1) (5 x+9) H_{-1}^2
                                +\frac{32\big(
                                        173 x^2+135 x+15\big)}{3} \zeta_2
                        \Biggr] H_0
        \nonumber\\ &&
                        +\Biggl[
                                \frac{2\big(
                                        2204 x^2+3190 x+1267\big)}{9} 
                                -16 \big(
                                        6 x^2+8 x+5\big) H_{-1}
                                -64 \big(
                                        2 x^2+3 x+3\big) \zeta_2
                        \Biggr] H_0^2
        \nonumber\\ &&
                        -\frac{8\big(
                                316 x^2+86 x+11\big)}{9}  H_0^3
                        +\frac{16}{3} (2 x+1) H_0^4
                        +\Biggl[
                                \frac{16}{9 x} \big(
                                        477 x^3-472 x^2+29 x-92\big) H_0
        \nonumber\\ &&
                                + \frac{16}{27 x} \big(
                                        2405 x^3-3223 x^2+923 x+371\big)
                                +\frac{16 \big(
                                        61 x^2-59 x-4\big)}{3 x} H_0^2
        \nonumber\\ &&
                                -16 \big(
                                        20 x^2-10 x-13\big) \zeta_2
                        \Biggr] H_1
                        +\Biggl[
                                - \frac{8}{3 x} \big(
                                        70 x^3-66 x^2+3 x-16\big) H_0
        \nonumber\\ &&
                                +\frac{16 \big(
                                        512 x^3-571 x^2+101 x-46\big)}{9 x} 
                        \Biggr] H_1^2
                        +\Biggl[
                                -32 (x+1) (18 x-37)
        \nonumber\\ && 
                                -64 (x+1) (5 x+9) H_{-1}
                                -32 \big(
                                        4 x^2-16 x+21\big) H_0
                                +16 \big(
                                        2 x^2-10 x-3\big) H_0^2
        \nonumber\\ && 
                                +128 (5 x+1) \zeta_2
                        \Biggr] H_{0,-1}
                        +32 \big(
                                6 x^2+2 x-1\big) H_{0,-1}^2
                        +\Biggl[
                                \frac{32}{3 x} \big(
                                        80 x^3-90 x^2+9 x-8\big) H_1
        \nonumber\\ &&  
                                - \frac{32}{3 x} \big(
                                        147 x^3+88 x^2-47 x-4\big) H_0
                                + \frac{16}{9 x} \big(
                                        405 x^3+197 x^2+143 x+92\big)
        \nonumber\\ &&                
                 +64 (x+1) (5 x+7) H_{-1}
                                -32 \big(
                                        9 x^2-2 x+7\big) H_0^2
                                +128 \big(
                                        5 x^2-7 x+4\big) \zeta_2
                        \Biggr] H_{0,1}
        \nonumber\\ &&                
         -64 \big(
                                2 x^2+3\big) H_{0,1}^2
                        +\Biggl[
                                64 (x+1) (5 x+9)
                                -128 \big(
                                        6 x^2+4 x+1\big) H_0
                        \Biggr] H_{0,-1,-1}
        \nonumber\\ &&                
                        +\Biggl[
                                -64 (x+1) (5 x+7)
                                +64 \big(
                                        2 x^2-10 x+1\big) H_0
                        \Biggr] H_{0,-1,1}
        \nonumber\\ &&                
          +\Biggl[
                                32 \big(
                                        14 x^2-24 x+47\big)
                                -32 \big(
                                        26 x^2-2 x+11\big) H_0
                        \Biggr] H_{0,0,-1}
                        +\Biggl[
                                \frac{32}{3 x} 
        \big(
                                        121 x^3+4 x^2
        \nonumber\\ &&               
        -50 x-4\big)
                                +32 \big(
                                        30 x^2+22 x+33\big) H_0
                        \Biggr] H_{0,0,1}
                        +\Biggl[
                                -64 (x+1) (5 x+7)
        \nonumber\\ &&                
                                +64 \big(
                                        2 x^2-10 x+1\big) H_0
                        \Biggr] H_{0,1,-1}
                        +\Biggl[
                                - \frac{32}{3 x} \big(
                                        197 x^3-85 x^2+11 x-8\big)
        \nonumber\\ &&               
                                -64 \big(
                                        8 x^2-14 x+1\big) H_0
                        \Biggr] H_{0,1,1}
                        +32 \big(
                                50 x^2+2 x+31\big) H_{0,0,0,-1}
                        -64 \big(
                                17 x^2+28 x+24\big) 
        \nonumber\\ &&  \times
        H_{0,0,0,1}
                        +64 \big(
                                2 x^2+4 x+5\big) H_{0,0,1,1}
                        -32 \big(
                                16 x^2-22 x+5\big) H_{0,1,1,1}
        \nonumber\\ &&                 
        -
                        \frac{16}{9} \big(
                                882 x^2+67 x+172\big) \zeta_2
                        -160 (x+1)^2 H_{-1}\zeta_2
                        +\frac{16}{5} \big(
                                40 x^2+202 x+73\big) \zeta_2^2
                \Biggr\}
        \nonumber\\ &&
                +\textcolor{blue}{C_F N_F^2 T_F^2 }
        \Biggl\{
                        - \frac{4}{81 x} \big(
                                36848 x^3-534 x^2-39075 x+2464\big)
                        + p_{qg}^{(0)}(x) \Biggl[
                                \frac{64}{3} H_0^2 H_1
                                -\frac{32}{9} H_1^3
        \nonumber\\ && 
                                -\frac{128}{3} H_0 H_{0,1}
                                +\frac{128}{3} H_{0,0,1}
                                -\frac{64}{3} H_{0,1,1}
                        \Biggr]
                        +\frac{16}{27} \big(
                                840 x^2+2635 x+2428\big) H_0
                        +\frac{8}{9} \big(
                                176 x^2
        \nonumber\\ && 
        +442 x+685\big) H_0^2
                        -\frac{32}{9} \big(
                                12 x^2+20 x-19\big) H_0^3
                        -\frac{32}{3} (2 x-1) H_0^4
                        +\Biggl[
                                -\frac{32}{27} \big(
        28 x^2-37 x
        \nonumber\\ &&                                
        +56\big)
                                +\frac{64}{9} \big(
                                        16 x^2-16 x+5\big) H_0
                        \Biggr] H_1
                        +\frac{64}{9} \big(
                                7 x^2-7 x+5\big) H_1^2
                        -\frac{64}{9} \big(
                                2 x^2-2 x-5\big) H_{0,1}
        \nonumber\\ && 
                        -\frac{128}{9} \big(
                                7 x^2-7 x+5\big) \zeta_2
                        +\frac{64}{3} p_{qg}^{(0)}(x)\zeta_3
        \Biggr\}
        \nonumber\\ &&
        +\textcolor{blue}{C_F^2 N_F T_F} \Biggl\{
                -963
                +1282 x
                -324 x^2
                + p_{qg}^{(0)}(-x) \Biggl[
                        -64 H_{-1}^2 H_0^2
                        +\frac{32}{3} H_{-1} H_0^3
        \nonumber\\ && 
                        +128 H_{-1} H_0 H_{0,-1}
                        -192 H_0 H_{0,0,-1}
                \Biggr]
                + p_{qg}^{(0)}(x) \Biggl[
                        \big(
                                -32 H_0^3
                                +448 H_0 \zeta_2
                        \big) H_1
        \nonumber\\ && 
                        +\big(
                                -64 H_0^2
                                +160 \zeta_2
                        \big) H_1^2
                        -
                        \frac{32}{3} H_0 H_1^3
                        -\frac{16}{3} H_1^4
                        +128 H_0 H_1 H_{0,-1}
        \nonumber\\ && 
                        +\big(
                                -128 H_0 H_1
                                -128 H_1^2
                                +128 H_{0,-1}
                        \big) H_{0,1}
                        -128 H_0 H_{0,-1,1}
                        -256 H_1 H_{0,0,-1}
        \nonumber\\ && 
                        +64 H_1 H_{0,0,1}
                        -128 H_0 H_{0,1,-1}
                        +192 H_1 H_{0,1,1}
                        -128 H_{0,-1,0,1}
                \Biggr]
                +\Biggl[
                        32 \big(
                                38 x^2-17 x+3\big)
        \nonumber\\ && 
                        +32 \big(
                                28 x^2-26 x+7\big) H_0
                        +448 p_{qg}^{(0)}(x) H_1
                \Biggr] \zeta_3
                +\Biggl[
                        -4 \big(
                                348 x^2+57 x+170\big)
        \nonumber\\ && 
                        +64 \big(
                                5 x^2+23 x+20\big) H_{-1}
                        -256 (x+1)^2 H_{-1}^2
                        +16 \big(
                                44 x^2-32 x+9\big) \zeta_2
                \Biggr] H_0
        \nonumber\\ && 
                +\Biggl[
                        -2 \big(
                                352 x^2+126 x+89\big)
                        +64 (x+1) (5 x+4) H_{-1}
                        +112 \big(
                                4 x^2-2 x+1\big) \zeta_2
                \Biggr] H_0^2
        \nonumber\\ && 
                -\frac{8}{3} \big(
                        60 x^2+16 x+3\big) H_0^3
                -\frac{8}{3} \big(
                        8 x^2+1\big) H_0^4
                +\Biggl[
                        -8 \big(
                                174 x^2-141 x-16\big)
        \nonumber\\ && 
                        -64 \big(
                                17 x^2-22 x+10\big) H_0
                        -32 (x-1) (5 x+2) H_0^2
                        +64 (x-1) (7 x+2) \zeta_2
                \Biggr] H_1
        \nonumber\\ && 
                +\Biggl[
                        -8 \big(
                                61 x^2-74 x+27\big)
                        -32 (x-1) (5 x+2) H_0
                \Biggr] H_1^2
                -\frac{8}{3} \big(
                        24 x^2-20 x-1\big) H_1^3
        \nonumber\\ && 
                +\Biggl[
                        -64 \big(
                                5 x^2+23 x+20\big)
                        +512 (x+1)^2 H_{-1}
                        -64 \big(
                                2 x^2+16 x-3\big) H_0
        \nonumber\\ && 
                        +32 \big(
                                6 x^2+2 x+3\big) H_0^2
                        -128 \zeta_2
                \Biggr] H_{0,-1}
                -64 \big(
                        2 x^2+2 x-1\big) H_{0,-1}^2
        +\Biggl[
                        16 \big(
                                7 x^2
        \nonumber\\ &&        
        +16 x+23\big)
                        +64 (2 x-5) H_0
                        -64 (x-1) (x+1) H_0^2
                        -128 (x-1) (3 x-2) H_1
        \nonumber\\ && 
                        -64 \big(
                                4 x^2-6 x+5\big) \zeta_2
                \Biggr] H_{0,1}
                -32 \big(
                        6 x^2-2 x+1\big) H_{0,1}^2
                +\big(
                        -512 (x+1)^2
                        -256 H_0
                \big) 
        \nonumber\\ &&   \times
        H_{0,-1,-1}
                -128 \big(
                        3 x^2-7 x+7\big) H_{0,0,-1}
                +\Biggl[
                        -16 \big(
                                24 x^2-12 x-23\big)
                        -64 \big(
                                2 x^2-2 x
        \nonumber\\ && 
        +5\big) H_0
                \Biggr] H_{0,0,1}
                +\Biggl[
                        16 \big(
                                44 x^2-66 x+15\big)
                        +128 \big(
                                7 x^2-6 x+3\big) H_0
                \Biggr] H_{0,1,1}
        \nonumber\\ && 
                -64 \big(
                        2 x^2-10 x+1\big) H_{0,0,0,-1}
                +32 \big(
                        8 x^2-10 x+17\big) H_{0,0,0,1}
                -64 \big(
                        8 x^2-10 x+5\big) H_{0,0,1,1}
        \nonumber\\ && 
                +32 \big(
                        8 x^2-6 x+3\big) H_{0,1,1,1}
                +16 \big(
                        61 x^2-12 x+17\big) \zeta_2
                -256 (x+1)^2 H_{-1} \zeta_2
        \nonumber\\ &&
                -
                \frac{128}{5} \big(
                        11 x^2-6 x+5\big) \zeta_2^2
        \Biggr\}
        \nonumber\\ &&
        +\textcolor{blue}{C_A^2 N_F T_F} 
        \Biggl\{
                - \frac{256}{3 x} (x+1) \big(
                        4 x^2+2 x+1\big) \left[ H_{0,-1,1} + H_{0,1,-1} \right]
        \nonumber\\ && 
                - \frac{8}{9 x} \big(
                        32 x^3-46 x^2+17 x+8\big) H_1^3
                - \frac{8}{27 x} \big(
                        51520 x^3-51304 x^2+5227 x-4702\big)
        \nonumber\\ &&         
        + p_{qg}^{(0)}(x) \Biggl[\Biggl(
                                -\frac{80}{3} H_0^2
                                +320  \zeta_2
                        \Biggr) H_1 H_0
                        +\big(
                                16 H_0^2
                                +32 \zeta_2
                        \big) H_1^2
                        +\frac{128}{3} H_0 H_1^3
                        -\frac{16}{3} H_1^4 
        \nonumber\\ && 
                        +192 H_0 H_1 H_{0,-1}
                        +\big(
                                64 H_0 H_1
                                -64 H_1^2
                                +192 H_{0,-1}
                        \big) H_{0,1}
                        -192 H_0 H_{0,-1,1}
                        -384 H_1 H_{0,0,-1}
        \nonumber\\ &&                
         -384 H_1 H_{0,0,1}
                        -192 H_0 H_{0,1,-1}
                        -192 H_{0,-1,0,1}
                \Biggr]
                + p_{qg}^{(0)}(-x) \Biggl[
                        -64 H_{-1}^3 H_0
                        -32 H_{-1}^2 H_0^2
        \nonumber\\ && 
                        +16 H_{-1} H_0^3
                        +\big(
                                192 H_{-1}^2
                                +192 H_{-1} H_0
                        \big) H_{0,-1}
                        -64 H_{-1}^2 H_{0,1}
                        -384 H_{-1} H_{0,-1,-1}
        \nonumber\\ && 
                        +128 H_{-1} H_{0,-1,1}
                        -256 H_{-1} H_{0,0,-1}
                        +128 H_{-1} H_{0,0,1}
                        +128 H_{-1} H_{0,1,-1}
                        -128 H_{-1} H_{0,1,1}
        \nonumber\\ && 
                        +384 H_{0,-1,-1,-1}
                        -128 H_{0,-1,-1,1}
                        -128 H_{0,-1,1,-1}
                        +128 H_{0,-1,1,1}
                        +256 H_{0,0,-1,-1}
        \nonumber\\ && 
                        -128 H_{0,0,-1,1}
                        -128 H_{0,0,1,-1}
                        -128 H_{0,1,-1,-1}
                        +128 H_{0,1,-1,1}
                        +128 H_{0,1,1,-1}
                        -32 H_{-1}^2 \zeta_2
                \Biggr]
        \nonumber\\ && 
                +
                \frac{32}{3 x} (x+1) \big(
                        8 x^2-41 x-4\big) \zeta_2 H_{-1}
                + \frac{16}{9 x} \big(
                        155 x^3-524 x^2+471 x-128\big) \zeta_2
        \nonumber\\ && 
                +\Biggl[
                        \frac{32}{3 x} \big(
                                346 x^3-375 x^2-18 x-36\big)
                        +160 p_{qg}^{(0)}(-x) H_{-1}
                        -768 (3 x+1) H_0
        \nonumber\\ && 
                        +992 p_{qg}^{(0)}(x) H_1
                \Biggr] \zeta_3
                +\Biggl[
                        \frac{32}{3 x} (x+1) \big(
                                40 x^2-25 x+4\big) H_{-1}^2
                        - \frac{32}{9 x} \big(
                                831 x^3+733 x^2
        \nonumber\\ && 
        -88 x+104\big) H_{-1}
                        +\frac{1}{x} \frac{8}{27} \big(
                                41040 x^3+15481 x^2+4348 x+448\big)
                        +\frac{16}{3} \big(
                                62 x^2-81 x+33\big) \zeta_2
                \Biggr]
        \nonumber\\ &&  \times
        H_0
                +\Biggl[
                        -\frac{8}{9} \big(
                                2561 x^2+450 x+599\big)
                        +\frac{16}{3 x} \big(
                                122 x^3+127 x^2+32 x+16\big) H_{-1}
        \nonumber\\ &&                
         +8 \big(
                                24 x^2-14 x+3\big) \zeta_2
                \Biggr] H_0^2
                -\frac{16}{9} \big(
                        36 x^2+7 x-7\big) H_0^3
                +\frac{8}{3} (15 x-4) H_0^4
        \nonumber\\ && 
                +\Biggl[
                        - \frac{8}{3 x} \big(
                                148 x^3-37 x^2-76 x-24\big) H_0^2
                        + \frac{16}{3 x} \big(
                                207 x^3-172 x^2-23 x-18\big) H_0
        \nonumber\\ && 
                        + \frac{8}{27 x} \big(
                                760 x^3+3875 x^2-4144 x-966\big)
                        + 
                        \frac{16}{3 x} \big(
                                34 x^3+16 x^2-53 x-8\big) \zeta_2
                \Biggr] H_1
        \nonumber\\ && 
                +\Biggl[
                        \frac{8}{3 x} \big(
                                126 x^3-98 x^2-23 x-16\big) H_0
                        - \frac{8}{9 x} \big(
                                423 x^3-476 x^2-41 x-40\big)
                \Biggr] H_1^2
        \nonumber\\ && 
                +\Biggl[
                        -\frac{64}{3 x} (x+1) \big(
                                40 x^2-25 x+4\big) H_{-1}
                        +\frac{32}{3 x} \big(
                                28 x^3-208 x^2-5 x-16\big) H_0
        \nonumber\\ &&                
         + \frac{32}{9 x} \big(
                                831 x^3+733 x^2-88 x+104\big)
                        -32 \big(
                                x^2-10 x+3\big) H_0^2
        \nonumber\\ &&                
         -32 \big(
                                8 x^2-2 x+7\big) \zeta_2
                \Biggr] H_{0,-1}
                +64 \big(
                        x^2-4 x+2\big) H_{0,-1}^2
        \nonumber\\ && 
                +\Biggl[
                        \frac{256}{3 x} (x+1) \big(
                                4 x^2+2 x+1\big) H_{-1}
                        -\frac{32}{3 x} \big(
                                40 x^3-26 x^2-17 x-8\big) H_1
        \nonumber\\ &&  
                        +\frac{16}{3 x} \big(
                                158 x^3-109 x^2-88 x-24\big) H_0
                        -\frac{32}{9 x} \big(
                                388 x^3+213 x^2+201 x+13\big)
        \nonumber\\ &&                
         +8 \big(
                                36 x^2+30 x+31\big) H_0^2
                        -32 \big(
                                20 x^2-10 x+13\big) \zeta_2
                \Biggr] H_{0,1}
                +32 \big(
                        10 x^2-2 x+7\big) H_{0,1}^2
        \nonumber\\ && 
                +\Biggl[
                        \frac{64}{3 x} (x+1) \big(
                                40 x^2-25 x+4\big)
                        -64 \big(
                                8 x^2-2 x+7\big) H_0
                \Biggr] H_{0,-1,-1}
        \nonumber\\ &&         
        +\Biggl[
                -\frac{32}{3 x} \big(
                                178 x^3-289 x^2+22 x-16\big)
                        -64 x (x+22) H_0
                \Biggr] H_{0,0,-1}
        \nonumber\\ &&
                +\Biggl[
                        - \frac{16}{3 x} \big(
                                230 x^3-100 x^2-67 x-24\big)
                        -64 \big(
                                13 x^2+20 x+13\big) H_0
                \Biggr] H_{0,0,1}
        \nonumber\\ &&         
        +\Biggl[
                        \frac{16}{x}  \big(
                                82 x^3+6 x^2-15 x-8\big)
                        -64 \big(
                                6 x^2+2 x+5\big) H_0
                \Biggr] H_{0,1,1}
                +96 \big(
                        2 x^2+22 x+5\big) 
        \nonumber\\ &&  \times
        H_{0,0,0,-1}
                +32 \big(
                        22 x^2+74 x+35\big) H_{0,0,0,1}
                +128 x (3 x-7) H_{0,0,1,1}
                +64 \big(
                        4 x^2-8 x+1\big) 
        \nonumber\\ &&  \times
        H_{0,1,1,1}
                -\frac{8}{5} \big(
                        112 x^2+990 x+269\big) \zeta_2^2
        \Biggr\}
        \nonumber\\ &&
        +\textcolor{blue}{C_A N_F^2 T_F^2} 
        \Biggl\{
                -\frac{16}{81 x} \big(
                        1400 x^3-2022 x^2+339 x-212\big)
                + p_{qg}^{(0)}(-x) \Biggl[
                        -\frac{64}{3} H_{-1} H_0^2
        \nonumber\\ && 
                        +\frac{128}{3} H_0 H_{0,-1}
                        -\frac{128}{3} H_{0,0,-1}
                \Biggr]
                + p_{qg}^{(0)}(x) \Biggl[
                        \frac{32}{3} H_0^2 H_1
                        +\frac{32}{3} H_0 H_1^2
                        +\frac{32}{9} H_1^3
                        +\Biggl(
                                -\frac{64}{3} H_0
        \nonumber\\ &&                       
          -\frac{128}{3} H_1
                        \Biggr) H_{0,1}
                        +\frac{64}{3} H_{0,0,1}
                        +64 H_{0,1,1}
                        +\frac{64}{3} H_1 \zeta_2
                \Biggr]
                +\frac{64}
                {27} \big(
                        229 x^2-70 x+20\big) H_0
        \nonumber\\ && 
                -\frac{128}{9} \big(
                        7 x^2+7 x+5\big) H_{-1} H_0
                -\frac{32}{9} \big(
                        8 x^2+39 x+8\big) H_0^2
                -\frac{64}{9} (4 x-1) H_0^3
        \nonumber\\ &&         
        +\frac{64}{27} \big(
                        14 x^2-14 x+19\big) H_1
                -\frac{64}{9} \big(
                        7 x^2-7 x+5\big) H_1^2
                +\frac{128}{9} \big(
                        7 x^2+7 x+5\big) H_{0,-1}
        \nonumber\\ &&
                -\frac{64}{3} x H_{0,1}
                -\frac{704 x \zeta_2}{9}
                +128 x \zeta_3
        \Biggr\}
        \Biggr\}
\end{eqnarray}
and
\begin{eqnarray}
P_{gg}^{(0)} &=&
        \left[\frac{22}{3} \textcolor{blue}{C_A} - \frac{8}{3} \textcolor{blue}{N_F T_F} \right] \delta(1-x)
        +\left(\frac{8 \textcolor{blue}{C_A}}{1-x}\right)_+
        - 8 \textcolor{blue}{C_A} \frac{\left(-1 + 2 x - x^2 +x^3\right)}{x},
\\
P_{gg}^{(1)} &=&
        -\left[
                \frac{32}{3}\textcolor{blue}{C_A N_F T_F}
                + 8 \textcolor{blue}{C_F N_F T_F}
                + \textcolor{blue}{C_A^2} \left(
                        -\frac{64}{3}
                        - 24 \zeta_3
                \right)
        \right] \delta(1-x)
        \nonumber\\ &&
        +\left(\frac{1}{1-x}\right)_+ \left[
                -\frac{160}{9} \textcolor{blue}{C_A N_F T_F}
                +\textcolor{blue}{C_A^2} \left[
                        -\frac{8}{9} \left(-67 + 18 \zeta_2\right)
                        +8 H_0^2 + 32 H_0 H_1
                \right]
        \right]
        \nonumber\\ &&  
        -\Biggl\{
                \textcolor{blue}{C_A^2} \Biggl[
                        \frac{4}{9} (109 x+25)
                        +\frac{8\big(2 x^3-2 x^2-4 x-1\big)}{x+1} H_0^2
                        -\frac{16 \big(2 x^3+2 x^2+4 x+3\big)}{x+1} \zeta_2
        \nonumber\\ &&
                        +\frac{8}{3} \big(44 x^2-11 x+25\big) H_0
                        -\frac{32 \big(x^2+x+1\big)^2}{x (x+1)} H_{-1} H_0
                        +\frac{32 \big(x^3-x^2+2 x-1\big)}{x} H_0 H_1
        \nonumber\\ &&
                        +\frac{32 \big(x^2+x+1\big)^2}{x (x+1)} H_{0,-1}
                \Biggr]
                +\textcolor{blue}{C_A N_F T_F} \Biggl[
                        -\frac{16}{9x} \big(23 x^3-19 x^2+29 x-23\big)
        \nonumber\\ &&
                        +\frac{32}{3} (x+1) H_0
                \Biggr]
                +\textcolor{blue}{C_F N_F T_F} \Biggl[
                        -\frac{32}{3x} (x-1) \big(5 x^2+11 x-1\big)
                        +16 (5 x+3) H_0
        \nonumber\\ &&
                        +16 (x+1) H_0^2
                \Biggr]
        \Biggr\},
\end{eqnarray}
\begin{eqnarray}
\lefteqn{P_{gg}^{(2),N_F} =}
\nonumber \\ &&
        \delta(1-x) \Biggl\{
                \textcolor{blue}{C_A^2 N_F T_F} \Biggl[
                        -\frac{466}{9}
                        -\frac{32}{3} \zeta_2^2
                        -\frac{320}{9} \zeta_3
                \Biggr]
                -\frac{482}{9} \textcolor{blue}{C_A C_F N_F T_F}
                +4 \textcolor{blue}{C_F^2 N_F T_F}
\nonumber \\ &&
                +\frac{116}{9} \textcolor{blue}{C_A N_F^2 T_F^2}
                +\frac{88}{9} \textcolor{blue}{C_F N_F^2 T_F^2}
        \Biggr\}
        +\Biggl(\frac{1}{1-x} \Biggl\{
                \textcolor{blue}{C_A^2 N_F T_F} \Biggl[
                        -\frac{32}{3} H_0
                        -\frac{320}{9} H_0^2
                        -\frac{128}{3} H_0 H_{0,-1}
\nonumber \\ &&
                        -\biggl(
                                \frac{1280}{9} H_0
                                +\frac{32}{3} H_0^2
                        \biggr) H_1
                        +\frac{64}{3} H_0 H_{0,1}
                        +\frac{256}{3} H_{0,0,-1}
                        -\frac{128}{3} H_{0,0,1}
                        -\frac{3344}{27}
                        +\frac{640}{9} \zeta_2
\nonumber \\ &&
                        -\frac{512}{3} \zeta_3
                \Biggr]
                +\textcolor{blue}{C_A C_F N_F T_F} \Biggl[
                        -\frac{440}{3}
                        +128 \zeta_3
                \Biggr]
                -\textcolor{blue}{C_A N_F^2 T_F^2} \frac{128}{27}
        \Biggr\}\Biggr)_+
\nonumber \\ &&
        +\textcolor{blue}{C_A^2 N_F T_F} \Biggl[
                -\frac{8}{81} \frac{9632 x^3-11037 x^2+9783 x-9632}{x}
                +\biggl(
                        32 \frac{\big(x+1\big) \big(2 x^2-9 x+2\big)}{x} H_{-1}^2
\nonumber \\ &&
                        +\frac{8}{27} \frac{3918 x^3-1979 x^2+3736 x+568}{x}
                        -\frac{32}{9} \frac{114 x^4-77 x^3-342 x^2-77 x+114}{\big(x+1\big) x} H_{-1}
                \biggr) H_0
\nonumber \\ &&
                +\biggl(
                        \frac{8}{9} \frac{124 x^3-481 x^2-657 x-92}{x+1}
                        -\frac{16}{3} \frac{\big(x+1\big) \big(4 x^2-31 x+4\big)}{x} H_{-1}
                \biggr) H_0^2
                +\frac{8}{3} x H_0^4
\nonumber \\ &&
                -\frac{16}{9} \big(19 x-12\big) H_0^3
                +\biggl(
                        -\frac{8}{27} \frac{\big(x-1\big) \big(410 x^2+2441 x+410\big)}{x}
                        -\frac{8}{3} \big(25 x-29\big) H_0^2
\nonumber \\ &&
                        +\frac{64}{9} \frac{33 x^3-29 x^2+49 x-33}{x} H_0
                \biggr) H_1
                +\biggl(
                        \frac{32}{9} \frac{114 x^4-77 x^3-342 x^2-77 x+114}{\big(x+1\big) x}
\nonumber \\ &&
                        -64 \frac{\big(x+1\big) \big(2 x^2-9 x+2\big)}{x} H_{-1}
                        -\frac{128}{3} \big(x^2+9 x-7\big) H_0
                        -96 \big(x-1\big) H_0^2
                \biggr) H_{0,-1}
\nonumber \\ &&
                +\biggl(
                        \frac{3664}{9} \big(x+1\big)
                        +\frac{128}{3} \frac{\big(x+1\big)^3}{x} H_{-1}
                        +\frac{16}{3} \big(17 x-37\big) H_0
                        +48 \big(x+1\big) H_0^2
                \biggr) H_{0,1}
\nonumber \\ &&
                -192 \big(x-1\big) H_{0,-1}^2
                +\biggl(
                        64 \frac{\big(x+1\big) \big(2 x^2-9 x+2\big)}{x}
                        +384 \big(x-1\big) H_0
                \biggr) H_{0,-1,-1}
\nonumber \\ &&
                -\frac{128}{3} \frac{\big(x+1\big)^3}{x} H_{0,-1,1}
                +\biggl(
                        \frac{32}{3} \frac{12 x^3+45 x^2-83 x+4}{x}
                        +64 \big(5 x-1\big) H_0
                \biggr) H_{0,0,-1}
\nonumber \\ &&
                -\biggl(
                        \frac{32}{3} \big(4 x^2+31 x-23\big)
                        +192 \big(x+1\big) H_0
                \biggr) H_{0,0,1}
                +384 \big(x+1\big) \big[H_{0,0,0,1} - H_{0,0,0,-1}\big]
\nonumber \\ &&
                -\frac{128}{3} \frac{\big(x+1\big)^3}{x} H_{0,1,-1}
                +\biggl(
                        \frac{32}{3} \frac{\big(x+1\big) \big(2 x^2-35 x+2\big)}{x} H_{-1}
                        +32 \frac{\big(x-1\big) \big(2 x^2+9 x+2\big)}{x} H_1
\nonumber \\ &&
                        +\frac{16}{3} \frac{8 x^4+43 x^3+34 x^2-5 x-8}{\big(x+1\big) x} H_0
                        -\frac{16}{9} \frac{132 x^4-137 x^3+156 x^2+481 x+96}{\big(x+1\big) x}
\nonumber \\ &&
                        -16 \big(5 x+3\big) H_0^2
                        +192 \big(x-1\big) H_{0,-1}
                        -192 \big(x+1\big) H_{0,1}
                \biggr) \zeta_2
                +\frac{48}{5} \big(13 x+23\big) \zeta_2^2
\nonumber \\ &&
                +\biggl(
                        \frac{64}{3} \frac{8 x^3+7 x^2+31 x-11}{x}
                        -384 x H_0
                \biggr) \zeta_3
        \Biggr]
        +\textcolor{blue}{C_A C_F N_F T_F} \Biggl[
                -\frac{16}{9} \big(70 x-29\big) H_0^3
\nonumber \\ &&
                -\frac{8}{81} \frac{12263 x^3-273 x^2-28806 x+15331}{x}
                +\biggl(
                        \frac{16}{27} \frac{756 x^3+3451 x^2+805 x-344}{x}
\nonumber \\ &&
                        -\frac{64}{9} \frac{\big(x+1\big) \big(10 x^2+275 x-44\big)}{x} H_{-1}
                        -\frac{64}{3} \frac{\big(x+1\big) \big(8 x^2-35 x+8\big)}{x} H_{-1}^2
                \biggr) H_0
\nonumber \\ &&
                -\frac{32}{3} \big(3 x-2\big) H_0^4
                +\biggl(
                        \frac{8}{9} \big(254 x^2+1208 x+1007\big)
                        -\frac{32}{3} \frac{\big(x+1\big) \big(2 x^2+25 x+2\big)}{x} H_{-1}
                \biggr) H_0^2
\nonumber \\ &&
                +\frac{32}{9} \frac{\big(x-1\big) \big(4 x^2+7 x+4\big)}{x} H_1^3
                +512 \big(x-1\big) H_{0,-1}^2
                +\biggl(
                        -\frac{16}{27} \frac{\big(x-1\big) \big(352 x^2-2219 x-521\big)}{x}
\nonumber \\ &&
                        +\frac{32}{9} \frac{\big(x-1\big) \big(37 x^2+79 x-17\big)}{x} H_0
                        +\frac{16}{3} \frac{\big(x-1\big) \big(16 x^2+55 x+16\big)}{x} H_0^2
                \biggr) H_1
\nonumber \\ &&
                +\biggl(
                        128 \big(x-1\big) H_0^2
                        +\frac{64}{9} \frac{\big(x+1\big) \big(10 x^2+275 x-44\big)}{x}
                        +\frac{128}{3} \frac{\big(x+1\big) \big(8 x^2-35 x+8\big)}{x} H_{-1}
\nonumber \\ &&
                        +\frac{128}{3} \frac{7 x^3+3 x^2-6 x+3}{x} H_0
                \biggr) H_{0,-1}
                +64 \big(x+1\big) H_{0,1}^2
                +\biggl(
                        -\frac{8}{9} \frac{\big(x-1\big) \big(4 x^2+31 x+4\big)}{x}
\nonumber \\ &&
                        +\frac{32}{3} \frac{\big(x-1\big) \big(4 x^2+7 x+4\big)}{x} H_0
                \biggr) H_1^2
                -\biggl(
                        \frac{16}{9} \frac{34 x^3+233 x^2+482 x+10}{x}
                        +\frac{256}{3} \frac{\big(x+1\big)^3}{x} H_{-1}
\nonumber \\ &&
                        +\frac{32}{3} \frac{12 x^3+38 x^2-37 x-8}{x} H_0
                        +224 \big(x+1\big) H_0^2
                        +\frac{64}{3} \frac{\big(x-1\big) \big(4 x^2+7 x+4\big)}{x} H_1
                \biggr) H_{0,1}
\nonumber \\ &&
                -\biggl(
                        \frac{128}{3} \frac{\big(x+1\big) \big(8 x^2-35 x+8\big)}{x}
                        +1024 \big(x-1\big) H_0
                \biggr) H_{0,-1,-1}
                +\frac{256}{3} \frac{\big(x+1\big)^3}{x} H_{0,-1,1}
\nonumber \\ &&
                -\biggl(
                        \frac{64}{3} \frac{26 x^3-15 x^2-51 x+10}{x}
                        +512 x H_0
                \biggr) H_{0,0,-1}
                +\frac{256}{3} \frac{\big(x+1\big)^3}{x} H_{0,1,-1}
                +\biggl(
                        640 \big(x+1\big) H_0
\nonumber \\ &&
                        +\frac{32}{3} \big(20 x^2+56 x-43\big)
                \biggr) H_{0,0,1}
                +\biggl(
                        \frac{32}{3} \frac{12 x^3+23 x^2+5 x-12}{x}
                        -128 \big(x+1\big) H_0
                \biggr) H_{0,1,1}
\nonumber \\ &&
                +768 \big(x+1\big) H_{0,0,0,-1}
                -960 \big(x+1\big) H_{0,0,0,1}
                -64 \big(x+1\big) H_{0,0,1,1}
                -128 \big(x+1\big) H_{0,1,1,1}
\nonumber \\ &&
                +\biggl(
                        512 \big(x+1\big) H_{0,1}
                        -\frac{16}{9} \frac{40 x^3+991 x^2-674 x-152}{x}
                        -\frac{64}{3} \frac{\big(x+1\big) \big(4 x^2-43 x+4\big)}{x} H_{-1}
\nonumber \\ &&
                        -\frac{32}{3} \frac{12 x^3+61 x^2-8 x-8}{x} H_0
                        -512 \big(x-1\big) H_{0,-1}
                        -\frac{64}{3} \frac{\big(x-1\big) \big(8 x^2+35 x+8\big)}{x} H_1
\nonumber \\ &&
                        +192 \big(x+1\big) H_0^2
                \biggr) \zeta_2
                -\frac{32}{5} \big(43 x+83\big) \zeta_2^2
                +\biggl(
                        \frac{32}{3} \frac{4 x^3-151 x^2-58 x+28}{x}
                        +64 \big(9 x-1\big) H_0
                \biggr) \zeta_3
        \Biggr]
\nonumber \\ &&
        +\textcolor{blue}{C_F^2 N_F T_F} \Biggl[
                \frac{8}{3} \frac{\big(x-1\big) \big(348 x^2+447 x+22\big)}{x}
                +\biggl(
                        \frac{8}{3} \big(108 x^2-541 x-381\big)
                        +\frac{4288}{3} \big(x+1\big) H_{-1}
\nonumber \\ &&
                        +\frac{256}{3} \frac{\big(x+1\big) \big(x^2-4 x+1\big)}{x} H_{-1}^2
                \biggr) H_0
                -\frac{16}{3} \big(x+1\big) H_0^4
                +\biggl(
                        -\frac{128}{3} \frac{\big(x+1\big) \big(x^2-4 x+1\big)}{x} H_{-1}
\nonumber \\ &&
                        +\frac{8}{3} \big(56 x^2-297 x-82\big)
                \biggr) H_0^2
                +\frac{16}{9} \big(4 x^2-33 x-9\big) H_0^3
                +\biggl(
                        \frac{16}{3} \frac{\big(x-1\big) \big(54 x^2-53 x-31\big)}{x}
\nonumber \\ &&
                        +\frac{32}{3} \frac{\big(x-1\big) \big(28 x^2+33 x+10\big)}{x} H_0
                        -\frac{16}{3} \frac{\big(x-1\big) \big(4 x^2+25 x+4\big)}{x} H_0^2
                \biggr) H_1
                -256 \big(x-1\big) H_{0,-1}^2
\nonumber \\ &&
                +\biggl(
                        \frac{8}{3} \frac{\big(x-1\big) \big(16 x^2+23 x+16\big)}{x}
                        -\frac{32}{3} \frac{\big(x-1\big) \big(4 x^2+7 x+4\big)}{x} H_0
                \biggr) H_1^2
                -\biggl(
                        128 \big(x-1\big) H_0^2
\nonumber \\ &&
                        +\frac{4288}{3} \big(x+1\big)
                        +\frac{512}{3} \frac{\big(x+1\big) \big(x^2-4 x+1\big)}{x} H_{-1}
                        +\frac{256}{3} \frac{\big(x-1\big) \big(x^2+4 x+1\big)}{x} H_0
                \biggr) H_{0,-1}
\nonumber \\ &&
                -\frac{32}{9} \frac{\big(x-1\big) \big(4 x^2+7 x+4\big)}{x} H_1^3
                +\biggl(
                        -\frac{16}{3} \frac{40 x^3-5 x^2-94 x-20}{x}
                        +96 \big(x+1\big) H_0^2
\nonumber \\ &&
                        -\frac{32}{3} \frac{4 x^3+27 x^2+51 x+4}{x} H_0
                        +\frac{64}{3} \frac{\big(x-1\big) \big(4 x^2+7 x+4\big)}{x} H_1
                \biggr) H_{0,1}
                -64 \big(x+1\big) H_{0,1}^2
\nonumber \\ &&
                +\biggl(
                        \frac{512}{3} \frac{\big(x+1\big) \big(x^2-4 x+1\big)}{x}
                        +512 \big(x-1\big) H_0
                \biggr) H_{0,-1,-1}
                +\biggl(
                        \frac{256}{3} \frac{3 x^3+3 x^2-9 x-1}{x}
\nonumber \\ &&
                        +512 x H_0
                \biggr) H_{0,0,-1}
                +\biggl(
                        \frac{32}{3} \frac{12 x^3+45 x^2+57 x+4}{x}
                        -384 \big(x+1\big) H_0
                \biggr) H_{0,0,1}
                +\biggl(
                        128 \big(x+1\big) H_0
\nonumber \\ &&
                        -\frac{64}{3} \frac{8 x^3+15 x^2+6 x-4}{x}
                \biggr) H_{0,1,1}
                -768 \big(x+1\big) H_{0,0,0,-1}
                +448 \big(x+1\big) H_{0,0,0,1}
\nonumber \\ &&
                +64 \big(x+1\big) H_{0,0,1,1}
                +128 \big(x+1\big) H_{0,1,1,1}
                +\biggl(
                        64 \big(x+1\big) H_0^2
                        -\frac{16}{3} \big(16 x^2-253 x+48\big)
\nonumber \\ &&
                        +\frac{256}{3} \frac{\big(x+1\big) \big(x^2-4 x+1\big)}{x} H_{-1}
                        +\frac{256}{3} \frac{\big(x-1\big) \big(x^2+4 x+1\big)}{x} H_1
                        +64 \big(5 x+4\big) H_0
\nonumber \\ &&
                        +256 \big(x-1\big) H_{0,-1}
                        -256 \big(x+1\big) H_{0,1}
                \biggr) \zeta_2
                +32 \big(9 x+13\big) \zeta_2^2
                -\biggl(
                        \frac{128}{3} \frac{8 x^3+3 x^2-2}{x}
\nonumber \\ &&
                        +64 \big(x-7\big) H_0
                \biggr) \zeta_3
        \Biggr]
        +\textcolor{blue}{C_A N_F^2 T_F^2} \Biggl[
                -\frac{16}{81} \frac{236 x^3-189 x^2+165 x-236}{x}
                -\frac{256}{9} \big(x+1\big) H_{0,1}
\nonumber \\ &&
                +\frac{16}{27} \big(52 x^2+43 x+76\big) H_0
                -\frac{64}{9} \big(x+1\big) H_0^2
                +\frac{16}{27} \frac{\big(x-1\big) \big(52 x^2+19 x+52\big)}{x} H_1
\nonumber \\ &&
                +\frac{256}{9} \big(x+1\big) \zeta_2
        \Biggr]
        +\textcolor{blue}{C_F N_F^2 T_F^2} \Biggl[
                -\frac{128}{81} \frac{\big(x-1\big) \big(85 x^2+49 x-77\big)}{x}
                +\frac{128}{27} \big(11 x^2+68 x-22\big) H_0
\nonumber \\ &&
                -\frac{64}{9} \big(x+1\big) H_0^3
                +\biggl(
                        \frac{64}{27} \frac{\big(x-1\big) \big(22 x^2+85 x-32\big)}{x}
                        +\frac{128}{9} \frac{\big(x-1\big) \big(4 x^2+7 x+4\big)}{x} H_0
                \biggr) H_1
\nonumber \\ &&
                +\frac{32}{9} \big(8 x^2+29 x+23\big) H_0^2
                +\frac{32}{9} \frac{\big(x-1\big) \big(4 x^2+7 x+4\big)}{x} H_1^2
                +\biggl(
                        -\frac{128}{9} \frac{2 x^3+7 x^2-2 x-4}{x}
\nonumber \\ &&
                        -\frac{256}{3} \big(x+1\big) H_0
                \biggr) H_{0,1}
                +\frac{256}{3} \big(x+1\big) H_{0,0,1}
                -\frac{128}{3} \big(x+1\big) H_{0,1,1}
                +\biggl(
                        -\frac{128}{9} \big(2 x^2-4 x-1\big)
\nonumber \\ &&
                        +\frac{256}{3} \big(x+1\big) H_0
                \biggr) \zeta_2
                -\frac{128}{3} \big(x+1\big) \zeta_3
        \Biggr].
\end{eqnarray}
\section{Calculation of the initial values}
\label{sec:B}

\vspace*{1mm}
\noindent
A sufficient number of initial values has to be provided to solve the recurrence relations which
determine the master integrals. These are the moments of the master integrals for
low values of the Mellin variable $N$. Up to $\mathcal{O}(\ep^0)$ they can be
calculated using the \texttt{MATAD} \cite{Steinhauser:2000ry} setup from
\cite{Bierenbaum:2009mv,Klein:2009ig}. Beyond this order, other methods have to
be used. If a multi-sum representation has already been derived to the necessary
order in $\ep$, the required initial values can be derived from it. However,
since this involves some degree of manual intervention and since an appropriate
representation cannot always be found, a fully algorithmic method is desirable.
In the following, we describe an algorithmic method which works for all master
integrals arising in the ongoing project to calculate the massive OMEs to three-loop order. 
The method has been briefly described in
\cite{Ablinger:2014uka} and applied in several calculations of this project
\cite{Ablinger:2014uka,Ablinger:2014nga,Ablinger:2015tua}.
It is based on the $\alpha$-parameterization for Feynman integrals, see, e.g., \cite{Itzykson:1980rh,
Smirnov:2006ry,Bogner:2010kv}, and the idea \cite{Tarasov:1996br,Lee:2009dh} of
rewriting integrals with scalar products in the numerator in terms of scalar
integrals in shifted dimensions with raised propagator powers. This allows us to
map fixed moments of the master integrals with operator insertions to
operator-less tadpole integrals. These can be reduced using integration-by-parts
(IBP) identities \cite{IBP,Studerus:2009ye,vonManteuffel:2012np} and lead to linear combinations of
only three
tadpole master integrals. Once these are calculated in shifted dimensions to
sufficient order in $\ep$, we can assemble the moments of the master integrals
with operator insertions. The steps described here can be implemented
efficiently in \texttt{Mathematica}.

\subsection{Structure of the integrals}
The scalar integrals with operator insertions for which we need initial values
are two-point integrals with an on-shell, massless external momentum $p^2=0$.
The Feynman rules of the operators involve a light-like vector $\Delta$ and they
introduce symbolic powers of scalar products $\Delta.k$, where $k$ is a
linear combination of loop momenta and $p$. Due to the structure of the Feynman
rules, we can deal with these scalar products elegantly by introducing an
ordinary generating function for the Mellin moments of the integrals. In the
simplest case of a scalar product $(\Delta.k)^N$, we introduce a tracing
variable $x$, multiply by $x^N$ and sum over $N$, cf.~Eq.~\eqref{eq:DPx}.
Similar terms appear also for more complicated operator insertions. The operator
insertions always lead to products of terms like
$(1 - x \, \Delta.k)^{-1}$, which resemble propagators that are linear in
the momentum $k$.

All scalar integrals of this project can be expressed in terms of 24 integral
families with nine quadratic propagators (which depend quadratically on a linear
combination of loop and external momenta) and three linear propagators that
encode the structure of the operator insertion. We use the notation
\begin{align}
  I^f(\nu_1,\dots,\nu_{12})
    &= \int \frac{\rmd^D k_1}{(2 \pi)^D} \frac{\rmd^D k_2}{(2 \pi)^D}
            \frac{\rmd^D k_3}{(2 \pi)^D}
            \prod_{i=1}^{12} P_i^{-\nu_i}
  \,,
\end{align}
where $P_i$ denotes the propagator denominators of integral family $f$. The
first nine indices refer to the quadratic propagators and the last three to the
linear propagators. The 24 twelve-propagator integral families are based on six
nine-propagator integral families (denoted B1, B3, B5, C1, C2 and C3), which are
supplemented by different sets of linear propagators (indicated by an additional
letter in the family identifier). For example, the integral family B5a has the
propagators
\begin{align*}
  P_{1} &= k_1^2-m^2                    \,, &
  P_{2} &= (k_1-p)^2-m^2                \,, &
  P_{3} &= k_2^2-m^2                    \,, \\
  P_{4} &= (k_2-p)^2-m^2                \,, &
  P_{5} &= k_3^2                        \,, &
  P_{6} &= (k_1-k_3)^2-m^2              \,, \\
  P_{7} &= (k_2-k_3)^2-m^2              \,, &
  P_{8} &= (k_1-k_2)^2                  \,, &
  P_{9} &= (k_3-p)^2                    \,, \\
  P_{10} &= 1-x \, \Delta.k_1     \,, &
  P_{11} &= 1-x \, \Delta.k_3     \,, &
  P_{12} &= 1-x \, \Delta.k_2     \,.
\end{align*}

We first discuss the properties of the quadratic propagators. The corresponding
nine-propagator families differ by the placement of the mass $m$ and by the
routing of the external momentum $p$. If no external linear propagator is
present, they can only depend on the mass $m$ and the external momentum $p$.
Thus, they reduce to massive tadpole integrals, since $p^2=0$, and we can remove
$p$ from the propagators altogether. Then three quadratic propagators become
linearly dependent of the remaining six ($P_2$, $P_4$ and $P_9$ in the example
above) and the six nine-propagator families reduce to only two six-propagator
tadpole integral families.

Let us briefly recapitulate the basic facts about the $\alpha$-parameterization
and its use for expressing integrals with irreducible numerators in terms of
scalar integrals in shifted dimensions. For more details, see for example
\cite{Itzykson:1980rh,Smirnov:2006ry,Bogner:2010kv,Tarasov:1996br,Lee:2009dh}.
The $\alpha$-parameterization for the above integrals without linear propagators
reads after applying a Wick rotation
\begin{align}
  I^f(\nu_1,\dots,\nu_9,0,0,0)
    &= \II^3 \int \frac{\rmd^D k_{1,E}}{(2 \pi)^D}
         \frac{\rmd^D k_{2,E}}{(2 \pi)^D} \frac{\rmd^D k_{3,E}}{(2 \pi)^D}
         \left[\prod_{\substack{1 \le i \le 9 \\ \nu_i > 0}}
               \int_0^\infty \rmd \alpha_i
               \frac{(-1)^{\nu_i} \alpha_i^{\nu_i-1}}{\Gamma(\nu_i)}\right]
\nonumber\\
& \times 
        \exp\left(-\sum_{\substack{1 \le i \le 9 \\ \nu_i > 0}}
                   \alpha_i P_{i,E}\right)
\end{align}
where we have explicitly pulled out a factor $(-1)$ from each propagator
$P_{i,E}$ due to the Wick rotation. Below we suppress the subscript $E$ and work
with Euclidean momenta unless explicitly stated otherwise.
Each propagator is at most quadratic in the loop momenta. Hence, we can
write any linear combination of propagators in the form
\begin{align}
  \sum_{\substack{1 \le i \le 9 \\ \nu_i > 0}} \alpha_i P_i
    &= \sum_{i,j=1}^3 k_i A_{ij} k_j + \sum_{i=1}^3 2 b_i k_i + c
  \,.
\end{align}
The matrix $A$ is real and symmetric, and the vectors $b_i$ will be linear
combinations of external momenta. Performing the loop integrals leads to
\begin{align}
  I^f(\nu_1,\dots,\nu_9,0,0,0)
    &= \frac{\II^3}{(2 \pi)^{3 D}}
         \left[\prod_{\substack{1 \le i \le 9 \\ \nu_i > 0}}
         \int_0^\infty \rmd \alpha_i
         \frac{(-1)^{\nu_i} \alpha_i^{\nu_i-1}}{\Gamma(\nu_i)}\right]
         \frac{\pi^{3 D/2}}{(\det A)^{D/2}}
         \exp\left(b^\T A^{-1} b - c\right)
  \,.
\end{align}
The elements of the matrix $A$ are the coefficients of the scalar products of
the loop momenta, while the elements of the vector $b$ are the coefficients of
the terms linear in the loop momenta -- i.e.\ scalar products with the external
momenta (or other vector quantities). The scalar term $c$ finally contains the
masses of the propagators. For our purposes, we have to consider the following
form of the propagators:
\begin{align}
  P_i &= \left(\sum_{j=1}^3 \delta^{(i)}_j k_j + \delta^{(i)}_p p\right)^2
         +\delta^{(i)}_m m^2
  \,.
\end{align}
The $\delta$ coefficients are either $1$, $0$ or $-1$ depending on how these
terms enter the propagator under consideration.
Each propagator enters the exponential multiplied by its corresponding $\alpha$
parameter if it is present in the integral. We can model this by replacing
$\alpha_i$ with $\beta_i = \alpha_i \Theta(\nu_i - \frac{1}{2})$, which vanishes
if the propagator is not present. Thus, we obtain
\begin{align}
  A_{ii} &=     \sum_{k=1}^9 \beta_k \delta^{(k)}_i                     \,, &
  A_{ij} &=     \sum_{k=1}^9 \beta_k \delta^{(k)}_i \delta^{(k)}_j      \,, &
  b_i    &= p   \sum_{k=1}^9 \beta_k \delta^{(k)}_i \delta^{(k)}_p      \,, &
  c      &= m^2 \sum_{k=1}^9 \beta_k \delta^{(k)}_m                     \,.
\end{align}
Here, we have already used that $p^2=0$.

For the integral families of this project, the explicit form of the matrix $A$
and the vector $b$ depend only on whether the integral belongs to one of the
families B1, B3 or B5 or to C1, C2 or C3. Using the notation introduced above,
the quantities read
\begin{align}
  &\begin{aligned}
    A^\text{B} &=
      \begin{pmatrix}
        \beta_1 + \beta_2 + \beta_6 + \beta_8     &
        -\beta_8                                  &
        -\beta_6                                  \\
        -\beta_8                                  &
        \beta_3 + \beta_4 + \beta_7 + \beta_8     &
        -\beta_7                                  \\
        -\beta_6                                  &
        -\beta_7                                  &
        \beta_5 + \beta_6 + \beta_7 + \beta_9     \\
      \end{pmatrix}
    \,,
  \end{aligned}
  \label{eq:AB} \\
  &\begin{aligned}
    A^\text{C} &=
      \begin{pmatrix}
        \beta_1 + \beta_2 + \beta_6 + \beta_8             &
        \beta_8                                           &
        -\beta_6 - \beta_8                                \\
        \beta_8                                           &
        \beta_3 + \beta_4 + \beta_7 + \beta_8             &
        -\beta_7 - \beta_8                                \\
        -\beta_6 - \beta_8                                &
        -\beta_7 - \beta_8                                &
        \beta_5 + \beta_6 + \beta_7 + \beta_8 + \beta_9   \\
      \end{pmatrix}
    \,,
  \end{aligned}
  \label{eq:AC} \\
  &\begin{aligned}
    b^\text{B} &= -p
      \begin{pmatrix}
        \beta_2 \\
        \beta_4 \\
        \beta_9 \\
      \end{pmatrix}
    \,, &
    b^\text{C} &= -p 
      \begin{pmatrix}
        \beta_2 + \beta_8 \\
        \beta_4 + \beta_8 \\
        \beta_9 - \beta_8\\
      \end{pmatrix}
    \,.
  \end{aligned}
  \label{eq:bBbC}
\end{align}
The scalar term $c$ is different for each of the six integral families. However,
we skip it here since its specific form does not play a role in the following.

\subsection{Treatment of the linear propagators}
If the integrals contain scalar products involving the loop momenta in the
numerator, we can include them in the $\alpha$-parameterization by introducing an
exponential generating function for them, i.e.
\begin{align}
  (q.k)^n &= \left.\left(\frac{1}{2} \frac{\rmd}{\rmd r}\right)^n
                     \exp(2 r \, q.k)\right|_{r=0}
  \label{eq:exp-genfunc}
  \,.
\end{align}
We introduce one auxiliary parameter $r$ for each scalar product. Thus,
auxiliary parameters associated to scalar products of two loop momenta, $k_i.k_j$,
end up as additional terms in the matrix elements $A_{ij}$ (e.g.,
$r_1 \, k_1.k_2$ in the exponential leads to the replacement $A_{12} \to
A_{12} + r_1$). The auxiliary parameters which accompany scalar products with
one loop momentum, $p.k_i$, are added to the parameter $b_i$ (e.g., the
scalar product $r_2 \, p.k_1$ yields $b_1 \to b_1 + r_2 p$).

To derive a parameter representation for fixed moments of the master integrals
with operator insertions, we first derive the $\alpha$-parameterization for the
corresponding operator-less integral. Afterwards, we add in the linear
propagators. Each of them has the structure $(1 - x \, \Delta.k)^{-1}$,
where $k$ is a linear combination of loop and external momenta. As discussed
above, it is introduced as an ordinary generating function for the moments of
the master integrals given in Eq.~\eqref{eq:DPx}.
If there are multiple linear propagators, all of them have the same tracing
parameter $x$, such that the expansion involves finite sums arising from the
Cauchy products. For example, two linear propagators yield
\begin{align}
  \frac{1}{1 - x \, \Delta.k_1} \frac{1}{1 - x \, \Delta.k_2}
    &= \sum_{N=0}^\infty x^N \sum_{j=0}^N (\Delta.k_1)^j
         (\Delta.k_2)^{N-j}
  \,,
\end{align}
which provides the generating function of the Feynman rule for an operator insertion on a vertex.
Analogously, for three linear propagators, we obtain
\begin{align}
  \frac{1}{1 - x \, \Delta.k_1} \frac{1}{1 - x \, \Delta.k_2}
    \frac{1}{1 - x \, \Delta.k_3}
    &= \sum_{N=0}^\infty x^N \sum_{k=0}^N \sum_{j=0}^k
         (\Delta.k_1)^j (\Delta.k_2)^{k-j} (\Delta.k_3)^{N-k}
  \,.
\end{align}
In case a linear propagator is raised to a higher power, we use
\begin{align}
  \frac{1}{(1 - x \, \Delta.k)^n} 
    &= \sum_{N=0}^\infty \binom{N+n-1}{n-1} x^{N} (\Delta.k)^{N}
  \,.
\end{align}
Thus, given a number of linear propagators, each of them raised to a given
power $n_i$, i.e.\ $(1 - x \, y_i)^{-n_i}$, we can extract the coefficient of
$x^N$ of the series expansion in $x$ as follows: We introduce a summation index
$j_i$ and a factor
\begin{align}
  \binom{(j_i-j_{i-1})+(n_i-1)}{n_i-1} y_i^{j_i-j_{i-1}}
\end{align}
for each linear propagator, multiply them together and sum over $j_i$ from
$0$ to $j_{i+1}$. Of course, there is no sum over $j_0$ and the outermost index
(which is never summed over) must be renamed to $N$.

To include the operators into our description, we think of the sums we just
introduced as a fully expanded polynomial in the scalar products of the linear
propagators. Each term has a number of scalar products each raised to a definite
power and for each term, we use the exponential generating function
representation, described in Eq.~\eqref{eq:exp-genfunc}. To summarize: For each
linear propagator we get a binomial coefficient  which encodes the
combinatorics of multiple, identical linear propagators, and we replace each
power of a scalar product by a corresponding power of the derivative operator
w.r.t. the auxiliary parameter $r_i$ (which belongs to the $i$th linear
propagator).

Given all these ingredients, we can assemble a general formula for the master
integrals with operator insertions for the case of three linear propagators
\begin{align}
I^f(\nu_1,\dots,\nu_{12})
    ={}& \frac{\II^3}{(4\pi)^{3 D/2}}
         \left[\prod_{\substack{1 \le i \le 9 \\ \nu_i > 0}}
         \int_0^\infty \rmd \alpha_i
         \frac{(-1)^{\nu_i} \alpha_i^{\nu_i-1}}{\Gamma(\nu_i)}\right]
  \nonumber \\ &
\vspace*{-4mm}
\times         \sum_{j_2=0}^{N} \sum_{j_1=0}^{j_2}
         \binom{j_1+(\nu_{10}-1)}{\nu_{10}-1}
         \binom{(j_2-j_1)+(\nu_{11}-1)}{\nu_{11}-1}
         \binom{(N-j_2)+(\nu_{12}-1)}{\nu_{12}-1}
  \nonumber \\ &
\vspace*{-4mm}
\times         \left(\frac{1}{2} \frac{\rmd}{\rmd r_1}\right)^{j_1}
         \left(\frac{1}{2} \frac{\rmd}{\rmd r_2}\right)^{j_2-j_1}
         \left(\frac{1}{2} \frac{\rmd}{\rmd r_3}\right)^{N-j_2}
         (\det A)^{-D/2}
         \exp(b^\T A^{-1} b - c)
         \Biggr|_{r_i = 0}
  \label{eq:master}
  \,.
\end{align}
Here, $A$ is completely determined by the operator-less part of the integral
($A^\text{B}$ and $A^\text{C}$ from Eqs.~\eqref{eq:AB}, \eqref{eq:AC}).
$b$ has one term, proportional to $p$, which is contributed by the operator-less
part ($b^\text{B}$ and $b^\text{C}$ from Eq.~\eqref{eq:bBbC}) and one part,
proportional to $r \Delta$, which stems from the linear
propagators. We describe this part in the next paragraph below. The scalar term
$c$ contains the mass term, but it remains unchanged by the operators, unless
the linear propagator involves the external momentum $p$.

The coefficients $b$ of the linear term are modified by the introduction of the
exponential generating functions for the operators: To each $b_i$ we have to add
a term $r_j \Delta$ where the linear propagators contain the loop momentum
$k_i$. Let us clarify this with an example: For family B1a we have the linear
propagators
\begin{align}
  P_{10} &= 1 - x \, \Delta.(k_3 - k_1)   \,, &
  P_{11} &= 1 - x \, \Delta.k_3           \,, &
  P_{12} &= 1 - x \, \Delta.(k_3 - k_2)   \,.
\end{align}
We introduce three parameters, $r_1$, $r_2$ and $r_3$, for the three
propagators and arrive at the exponential generating function
\begin{align}
  \exp\left(2 r_1 \Delta.(k_3 - k_1)
           +2 r_2 \Delta.k_3
           +2 r_3 \Delta.(k_3-k_2)\right)
  \,,
\end{align}
and thus we obtain for $b$
\begin{align}
  b^\text{B1a} &= b^\text{B} + \Delta
    \begin{pmatrix}
      - r_1 \\
      - r_3 \\
      r_1 + r_2 + r_3
    \end{pmatrix}
  \,.
\end{align}
Similar terms arise for the other integral families.

If the linear propagators contain the external momentum $p$, we have to also
modify the scalar term $c$. In that case we have to add a term proportional to
$\Delta.p$. As an example, we use family B3b, which has the linear
propagators
\begin{align}
  P_{10} &= 1 - x \, \Delta.(k_1 - p)     \,, &
  P_{11} &= 1 - x \, \Delta.k_3           \,, &
  P_{12} &= 1 - x \, \Delta.(k_2 - p)     \,,
\end{align}
and leads  to the exponential generating function
\begin{align}
  \exp(2 r_1 \Delta.(k_1 - p)
      +2 r_2 \Delta.k_3
      +2 r_3 \Delta.(k_2 - p))
\end{align}
and hence
\begin{align}
  b^\text{B3b} &= b^\text{B} + \Delta
    \begin{pmatrix}
      r_1 \\
      r_3 \\
      r_2
    \end{pmatrix}
  \,, &
  c^\text{B3b} &= c^\text{B3} + 2 \Delta.p (-r_1-r_3)
  \,.
\end{align}

The central idea of the dimensional shifts is now that derivatives with respect
to $r_i$ will act on the exponential function and introduce factors of
$(A^{-1})_{ij} = (\det(A))^{-1} \tilde{A}_{ij} $, where $\tilde{A}_{ij}$ is the
$(i,j)$ cofactor of the matrix $A$. Since the dimension $D$ enters the
parametric representation Eq.~\eqref{eq:master} only in the exponent of the
determinant (except for constant prefactors), we can reinterpret the integrals
arising from taking the derivatives as integrals in shifted dimensions,
according to the exponent of the determinant. The cofactors $\tilde{A}_{ij}$ are
polynomials in the $\alpha$ parameters and can be absorbed into the powers of
$(\alpha_k)^{\nu_k-1}$ which encode the powers to which the quadratic
propagators are raised. More concretely, the powers of the propagators enter the
$\alpha$ representation in the form
\begin{align}
  \prod_{i=1}^n \frac{(-1)^{\nu_i} \alpha_i^{\nu_i-1}}{\Gamma(\nu_i)}
\end{align}
which means that an additional factor of $\alpha_j^k$ in the integrand
corresponds to raising the power of propagator $P_j$ by $k$ (i.e., $\nu'_j =
\nu_j + k$) and multiplying the expression by, cf. e.g. \cite{SLATER},
\begin{align}
  (-1)^k \frac{\Gamma(\nu_i+k)}{\Gamma(\nu_i)} = (-1)^k (\nu_i)_k
\end{align}
to compensate for the changed $(-1)^{\nu'_i}$ and $\Gamma(\nu'_i)$.
Here $(\nu_i)_k$ denotes the Pochhammer symbol.
Thus, each derivative shifts the dimension and raises the powers of some
propagators. After applying all derivatives, we obtain a linear combination of
scalar integrals in shifted dimensions with dotted propagators and no operator
insertions. These can then be mapped to tadpole integrals.

Let us now have a closer look at how the derivatives act on the parameter
representation. For the master integrals we are concerned with, the auxiliary
parameters $r_i$ only enter $b$ and $c$.
The vector $b$ has two terms---a term proportional to $p$ and one proportional
to $\Delta$. Since $b$ only appears in the bilinear form $b^\T A^{-1} b$ and
since $p^2=0$ and $\Delta^2=0$, only the mixed terms proportional to $\Delta.p$
survive. Therefore, all terms are linear in the $r_i$. Thus, each
derivative acting on this part of the exponential brings down exactly one matrix
element of $A^{-1}$ and one power of $\Delta.p$. This means that each
derivative shifts the dimension of the integral from $D$ to $D+2$.
For the scalar term $c$ we also have a linear dependence on the $r_i$. Here,
however, we do not get any shift of dimension from a derivative acting on this
part of the exponential, since it only brings down a power of
$2 \Delta.p$. This leads to linear combinations of integrals with
different dimensional shifts. In all cases, each derivative introduces a factor
$\Delta.p$, so that the $N$th moment of any master integral will always be
proportional to $(\Delta.p)^N$.

Finally, we have to undo the Wick rotation. The only scalar product left in the
final expression is $(\Delta.p)^N$ which introduces an additional, global
factor of $(-1)^N$ into \eqref{eq:master}.

\subsection{Reduction of the tadpole integrals}
\begin{figure}
  \centering
  \subfloat[$I^\text{B1a}_{112}$]{%
    \includegraphics[scale=0.8]{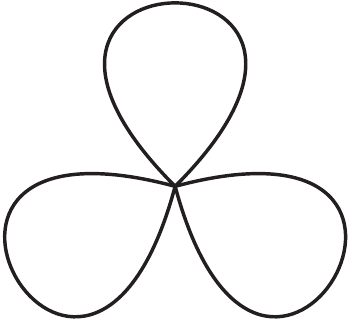}} \hfill
  \subfloat[$I^\text{B1a}_{101}$]{%
    \includegraphics[scale=0.8]{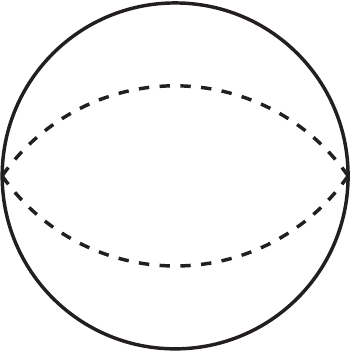}} \hfill
  \subfloat[$I^\text{B5a}_{101}$]{%
    \includegraphics[scale=0.8]{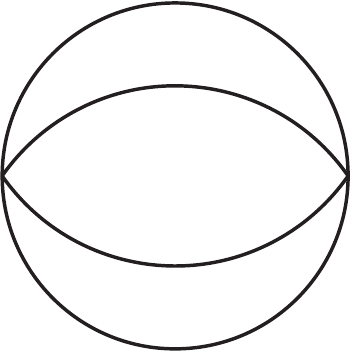}}
  \caption{\sf Diagrams for the tadpole master integrals. Dashed lines represent
           massless propagators and massive propagators are depicted as solid
           lines.}
  \label{fig:tadpoles}
\end{figure}
The procedure described above maps a scalar integral with an operator insertion
in $D=4+\ep$ dimensions to a (possibly large) linear combination of
operator-less tadpole integrals in shifted dimensions. To cope with these, we
use IBP relations derived with \texttt{Reduze 2} to map them onto linear
combinations of master integrals. Their coefficients are rational functions in
the dimension $D$ and the mass squared $m^2$. The reductions are done
for general values of $D$ so that they can be used for integrals in any
dimension. Particular care must be taken if the initial values are mapped to a
linear combination of tadpole integrals in different dimensions.

As already observed in \cite{Broadhurst:1991fi}, all tadpole
integrals can be reduced to just three master integrals. They correspond to
\begin{align}
  I^\text{B1a}_{112} &\coloneqq I^\text{B1a}(0,0,0,0,1,1,1,0,0,0,0,0)  \,, \\
  I^\text{B1a}_{101} &\coloneqq I^\text{B1a}(1,0,1,0,0,1,1,0,0,0,0,0)  \,, \\
  I^\text{B5a}_{101} &\coloneqq I^\text{B5a}(1,0,1,0,0,1,1,0,0,0,0,0)  \,,
\end{align}
and their diagrams are shown in Figure~\ref{fig:tadpoles}.

\subsection{Required tadpole integrals}
The first two tadpole integrals are straightforwardly obtained and can be expressed
completely in terms of $\Gamma$-functions, for example using their Feynman
parameterization. The first one is just the one-loop tadpole cubed, which reads
\begin{align}
  I^\text{B1a}_{112}
    &= \int \frac{\rmd^D k_1}{(2 \pi)^D} \frac{\rmd^D k_2}{(2 \pi)^D}
            \frac{\rmd^D k_3}{(2 \pi)^D} \frac{1}{k_3^2 - m^2}
            \frac{1}{(k_1-k_3)^2 - m^2} \frac{1}{(k_2-k_3)^2-m^2}
  \nonumber \\
    &= -\mathcal{N} m^{3 D - 6} \Gamma \left(1-\frac{D}{2}\right)^3
  \,,
\end{align}
where we use the normalization factor
\begin{align}
  \mathcal{N} &= \left({\rm i} (4 \pi)^{-D/2}\right)^3
  \,,
\end{align}
and the second one is the three-loop banana graph with two massless lines,
yielding
\begin{align}
  I^\text{B1a}_{101}
    &= \int \frac{\rmd^D k_1}{(2 \pi)^D} \frac{\rmd^D k_2}{(2 \pi)^D}
            \frac{\rmd^D k_3}{(2 \pi)^D} \frac{1}{k_1^2} \frac{1}{k_2^2}
            \frac{1}{(k_1-k_3)^2 - m^2} \frac{1}{(k_2-k_3)^2 - m^2}
  \nonumber \\
    &= \mathcal{N} m^{3 D - 8}
       \frac{\Gamma\left(4-\frac{3 D}{2}\right) \Gamma\left(3-D\right)^2
             \Gamma\left(\frac{D}{2}-1\right)^2
             \Gamma\left(2-\frac{D}{2}\right)}{%
             \Gamma\left(6-2 D\right) \Gamma\left(\frac{D}{2}\right)}
  \,.
\end{align}
These expressions can be expanded in $\ep$ in any dimension $D = 2 n + \ep$.
When expanding in $\ep$, we keep the normalization $\mathcal{N}$, the mass scale
$m^{3 D - 2\nu}$ and an additional factor of $\exp(\frac{3}{2} \ep \EulerGamma)$
unexpanded to simplify the expressions. The latter is the same regardless of the
integer dimension around which we expand.
The third integral is the three-loop banana graph with fully massive lines and
reads
\begin{align}
  I^\text{B5a}_{101}
    &= \int \frac{\rmd^D k_1}{(2 \pi)^D} \frac{\rmd^D k_2}{(2 \pi)^D}
            \frac{\rmd^D k_3}{(2 \pi)^D}
            \frac{1}{k_1^2 - m^2} \frac{1}{k_2^2 - m^2}
            \frac{1}{(k_1-k_3)^2 - m^2} \frac{1}{(k_2-k_3)^2 - m^2}.
\end{align}
In \cite{Broadhurst:1991fi} this integral was found to correspond to an
expression involving $\Gamma$-functions and a generalized hypergeometric function
${}_3F_2$ for general values of $D$. Here, we are interested in the expansion
around $\ep=D-2n$ for different values of $n \in \mathbb{N}$ up to $2n = 30$. To
calculate it, we follow the method used in \cite{Bekavac:2009gz,Grigo:2012ji,
Schroeder:LL2016}, which we briefly describe below.

\begin{figure}
  \centering
  \subfloat[\label{subfig:J1} $J_1(x)$]{%
    \includegraphics[scale=0.8]{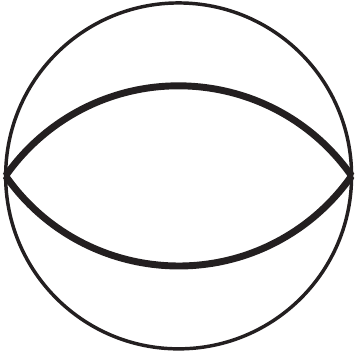}} \hfill
  \subfloat[\label{subfig:J2} $J_2(x)$]{%
    \includegraphics[scale=0.8]{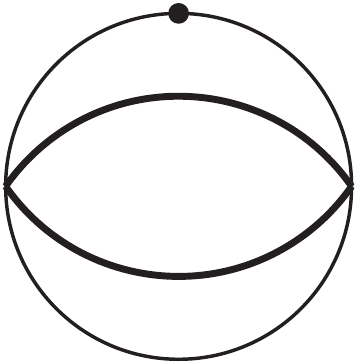}} \hfill
  \subfloat[\label{subfig:T1} $T_1(x)$]{%
    \includegraphics[scale=0.8]{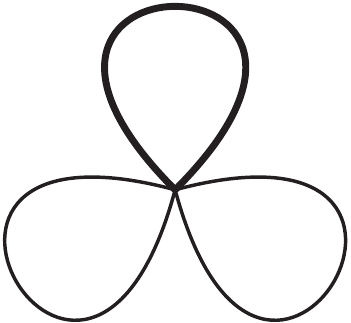}} \hfill
  \subfloat[\label{subfig:T2} $T_2(x)$]{%
    \includegraphics[scale=0.8]{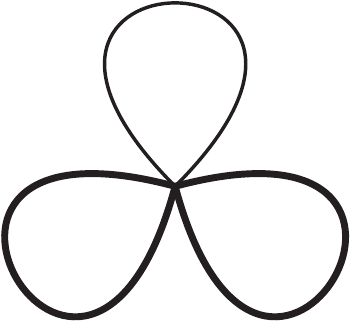}}
  \caption{\sf Diagrams for the master integrals with propagators of two different
           masses. Thin solid lines denote propagators with mass $m_1$ and thick
           solid lines correspond to propagators with mass $m_2$.}
  \label{fig:tadpoles-2mass}
\end{figure}
We calculate this integral as a special case of the same integral with two
different masses, by deriving a differential equation in the ratio of the masses
and then specializing to the case of equal masses. The two-mass diagram is
depicted in Figure~\ref{subfig:J1}.
\begin{align}
  J_a(x)
    &= \int \frac{\rmd^D k_1}{(2 \pi)^D} \frac{\rmd^D k_2}{(2 \pi)^D}
            \frac{\rmd^D k_3}{(2 \pi)^D}
            \frac{1}{(k_1^2 - m_2^2)^a} \frac{1}{k_2^2 - m_2^2}
            \frac{1}{(k_1-k_3)^2 - m_1^2} \frac{1}{(k_2-k_3)^2 - m_1^2}
  \nonumber \\
    &= m_1^{3 D - 6 - 2 a}
       \int \frac{\rmd k_1}{(2 \pi)^D} \frac{\rmd k_2}{(2 \pi)^D}
            \frac{\rmd k_3}{(2 \pi)^D}
            \frac{1}{(k_1^2 - x^2)^a} \frac{1}{k_2^2 - x^2}
            \frac{1}{(k_1-k_3)^2 - 1} \frac{1}{(k_2-k_3)^2 - 1}
  \,,
\end{align}
where we define $x = \frac{m_2}{m_1}$. Since the overall factor of $m_1^\alpha$,
$\alpha \in \mathbb{R}$ can be reconstructed from dimensional arguments, we set
$m_1=1$ in the remaining discussion. For $x=0$ the integral with $a=1$
corresponds to $J_1(0) = I^\text{B1a}_{101}$ while for $x=1$ it corresponds to
$J_1(1) = I^\text{B5a}_{101}$. Moreover, we can derive the property
\begin{align}
  J_1(x) &= x^{3 D-8} J_1(x^{-1})
  \label{eq:fliprel}
\end{align}
from the unevaluated integral.

The system of differential equations with respect to $x$ can be derived using
for example \texttt{Reduze 2} \cite{Studerus:2009ye,vonManteuffel:2012np}.
In order to close the system, four integrals are necessary. Besides $J_1(x)$,
also $J_2(x)$, i.e.\ the same integral with the first propagator raised to the
second power (see Figure~\ref{subfig:J2}), contributes. Moreover, the integrals
$T_1(x)$ and $T_2(x)$ appear (see Figure~\ref{subfig:T1} and \ref{subfig:T2},
respectively). They correspond to products of the massive one-loop tadpole
integral and can be expressed in terms of $\Gamma$-functions for any value of $D$.
Expressed in terms of the single scale tadpole they evaluate to (again setting
$m_1=1$)
\begin{align}
  T_1(x) &= x^{D/2-1} I^\text{B1a}_{101}        \,, &
  T_2(x) &= x^{D-2} I^\text{B1a}_{101}		\,.
\end{align}
Therefore, we treat them as known inhomogeneous parts of the differential
equation. The resulting system of first order differential equations reads
\begin{align*}
  \frac{\rmd J_1(x)}{\rmd x} ={}& 4x J_2(x)
  \,, \\
  \frac{\rmd J_2(x)}{\rmd x} ={}&
    \frac{(D-3) (3D-8)}{4 x (1-x) (1+x)} J_1(x)
    +\frac{2D-7 + x^2 (13-4D)}{x (1-x) (1+x)} J_2(x)
  \nonumber \\ & 
    +\frac{(D-2)^2}{4 x^3 (1-x) (1+x)}
      \left[T_1(x) - T_2(x) \right]
  \,.
\end{align*}
These equations can be uncoupled into a single, second order differential
equation -- either by using the package \texttt{OreSys}
\cite{UNCOUPL,Gerhold:2002} or here even by hand. We obtain
\begin{align*}
  &\frac{\rmd^2 J_1(x)}{\rmd x^2}
    -\frac{2 (D-3) (1-2x^2)}{x (1-x) (1+x)} \frac{\rmd J_1(x)}{\rmd x}
    -\frac{(3D-8) (D-3)}{(1-x) (1+x)} J_1(x)
  \nonumber \\
  &\qquad = \frac{(D-2)^2}{x^2 (1-x) (1+x)} \left[T_1(x) - T_2(x)\right]
  \,,
\end{align*}
which is a second order differential equation with regular singular points at
$x \in \{-1,0,1,\infty\}$. Thus, we can only use the initial condition
$J_1(0)=I^\text{B1a}_{101}$, while the derivative of $J_1$ at $x=0$ becomes
linearly dependent. We could use the value of $J_1$ at $x=1$ to fix the second
constant, however, this is just the quantity we want to calculate here. Therefore,
we use the relation between $J_1(x)$ and $J_1(x^{-1})$ given in
Eq.~\eqref{eq:fliprel} instead.

The differential equation can be expanded in $\ep=D-2 n$, $n \in \mathbb{N}$ and
solved order by order. The homogeneous equation is the same at every order and
therefore also the homogeneous solutions are the same. They can be found for
example using the \texttt{Mathematica} routine \texttt{DSolve}. The solutions contain
polynomials and logarithms in $x$. For $D=4$ they are given by
\begin{align}
  h_1(x) &= (1-x)^2 (1+x)^2
  \\
  h_2(x) &= \frac{x \big(x^2+1\big)}{8}
            -\frac{(1-x)^2 (1+x)^2}{16} \bigl[H_{-1}(x) + H_1(x)\bigr]
  \,.
\end{align}
The inhomogeneous part $r(x)$ of the differential equation arises from the
expansion of $T_1(x)$ and $T_2(x)$ in $\ep$ as well as lower orders of the
solution to $J_1(x)$. We then use the variation of constants to calculate a
particular solution to the inhomogeneous differential equation, cf. \cite{KAMKE},
\begin{align}
  p(x) &= -h_1(x) \int_0^x \frac{r(x')}{W(x')} h_2(x') \, \rmd x'
          +h_2(x) \int_0^x \frac{r(x')}{W(x')} h_1(x') \, \rmd x'
  \,,
\end{align}
where $W(x) = h_1(x) h_2'(x) - h_2(x) h_1'(x)$ is the Wronskian of the
homogeneous solutions. For $D=4$ it is $W(x) = x^2 (1-x) (1+x)$. The structure
of these integrals tells us that they only lead to iterated integrals over the
letters $x$, $1-x$ and $1+x$, i.e.\ the harmonic polylogarithms \cite{Remiddi:1999ew}.
Performing these integrals can be done efficiently using the \texttt{HIntegrate}
command from the package \texttt{HarmonicSums}.
Finally, we fix the constants of integration using the expansion of
$J_1(0) = I^\text{B1a}_{112}$ along with the relation Eq.~\eqref{eq:fliprel}.
Since only HPLs occur, the relation maps them to linear combinations of HPLs
again. If both sides are algebraically reduced, we can solve for the constants
by comparing coefficients of the HPLs. Of course, this requires the algebraic
relations for HPLs of argument $x$ to be available. In intermediate steps of the
calculation at hand, we needed these relations up to weight ${\sf w = 9}$, which can be
calculated, e.g., using \texttt{HarmonicSums}, cf. also \cite{Vermaseren:1998uu,Blumlein:2009cf}.
The expansion of the solution for
$J_1(x)$ is given completely in terms of HPLs with polynomial coefficients in
$x$. For $D=4+\ep$ we obtain
\begin{align}%
        \MoveEqLeft{J_1(x) =}
\nonumber \\ &
        \mathcal{N} \E^{\frac{3}{2} \ep \EulerGamma} \biggr\{
        -\frac{1}{\ep^3} \frac{8}{3} \big(x^4+4 x^2+1\big)
        +\frac{1}{\ep^2} \biggl[
                \frac{2}{3} \big(7 x^4+32 x^2+7\big)
                -8 \big(x^2+2\big) x^2 H_0
        \biggr]
\nonumber \\ &
        +\frac{1}{\ep} \biggl[
                2 \big(7 x^2+16\big) x^2 H_0
                -\frac{5}{6} \big(5 x^4+32 x^2+5\big)
                -\big(x^4+4 x^2+1\big) \zeta_2
                -4 \big(3 x^2+2\big) x^2 H_0^2
        \biggr]
\nonumber \\ &
        -\frac{5}{24} \big(x^4-112 x^2+1\big)
        -\frac{5}{2} \big(5 x^2+16\big) x^2 H_0
        +3 \big(7 x^2+4\big) x^2 H_0^2
        -\frac{4}{3} \big(7 x^2+2\big) x^2 H_0^3
\nonumber \\ &
        +\biggl(
                \frac{1}{4} \big(7 x^4+32 x^2+7\big)
                -3 \big(x^2+2\big) x^2 H_0
        \biggr) \zeta_2
        +\frac{1}{3} \big(7 x^4-20 x^2+7\big) \zeta_3
        +\bigl[
                4 H_0^2 H_1
\nonumber \\ &
                -4 H_{-1} H_0^2
                +8 H_0 H_{0,-1}
                -8 H_0 H_{0,1}
                -8 H_{0,0,-1}
                +8 H_{0,0,1}
        \bigr] (1-x)^2 (1+x)^2
        +\mathcal{O}(\ep)
        \biggr\},
\end{align}%

\noindent
where we suppressed the higher order terms in $\ep$ due to their length.

To obtain the solution for $I^\text{B5a}_{101} = J_1(1)$, we express the HPLs at
$x = 1$ in terms of multiple zeta values and use their relations
\cite{Blumlein:2009cf} to reduce the occurring constants to an algebraically
independent basis. The full result for $D=4+\ep$ reads
\begin{align}%
        \MoveEqLeft{I^\text{B5a}_{101} =}
\nonumber \\ &
        \mathcal{N} \E^{\frac{3}{2} \ep \EulerGamma} \Biggl\{
        -\frac{16}{\ep^3}
        +\frac{92}{3 \ep^2}
        -\frac{
                35
                +6 \zeta_2
        }{\ep}
        +\frac{275}{12}
        +\frac{23}{2} \zeta_2
        -2 \zeta_3
        +\ep \biggl[
                \frac{189}{16}
                -\frac{105}{8} \zeta_2
                -\frac{89}{6} \zeta_3
                -\frac{57}{16} \zeta_4
        \biggr]
\nonumber \\ &
        +\ep^2 \biggl[
                -\frac{14917}{192}
                +\frac{275}{32} \zeta_2
                +\frac{525}{8} \zeta_3
                -\frac{3}{4} \zeta_2 \zeta_3
                -\frac{2251}{64} \zeta_4
                -\frac{3}{10} \zeta_5
                +4 B_4
        \biggr]
        +\ep^3 \biggl[
                \frac{48005}{256}
\nonumber \\ &
                +\frac{567}{128} \zeta_2
                -\frac{15965}{96} \zeta_3
                -\frac{89}{16} \zeta_2 \zeta_3
                -\frac{1}{8} \zeta_3^2
                +\frac{38325}{256} \zeta_4
                +\frac{6223}{40} \zeta_5
                -\frac{631}{512} \zeta_6
                -15 B_4
\nonumber \\ &
                -192 \text{Li}_5\left(\frac{1}{2}\right)
                -204 \zeta_4 \ln(2)
                -16 \zeta_2 \ln(2)^3
                +\frac{8}{5} \ln(2)^5
        \biggr]
        +\ep^4 \biggl[
                -\frac{1108525}{3072}
                -\frac{14917}{512} \zeta_2
\nonumber \\ &
                +\frac{43309}{128} \zeta_3
                +\frac{1575}{64} \zeta_2 \zeta_3
                -\frac{9961}{96} \zeta_3^2
                -\frac{384535}{1024} \zeta_4
                +\frac{39}{4} \zeta_2 \zeta_4
                -\frac{57}{128} \zeta_3 \zeta_4
                -\frac{18621}{32} \zeta_5
\nonumber \\ &
                -\frac{9}{80} \zeta_2 \zeta_5
                -\frac{2582095}{6144} \zeta_6
                -\frac{3}{56} \zeta_7
                +\biggl(
                        \frac{145}{4}
                        +\frac{3}{2} \zeta_2
                \biggr) B_4
                +720 \text{Li}_5\left(\frac{1}{2}\right)
                +576 \text{Li}_6\left(\frac{1}{2}\right)
\nonumber \\ &
                +765 \zeta_4 \ln(2)
                -306 \zeta_4 \ln(2)^2
                +60 \zeta_2 \ln(2)^3
                -12 \zeta_2 \ln(2)^4
                -6 \ln(2)^5
                +\frac{4}{5} \ln(2)^6
\nonumber \\ &
                +240 s_6
        \biggr]
        +\ep^5 \biggl[
                \frac{2570029}{4096}
                +\frac{144015}{2048} \zeta_2
                -\frac{942389}{1536} \zeta_3
                -\frac{15965}{256} \zeta_2 \zeta_3
                +\frac{49885}{128} \zeta_3^2
\nonumber \\ &
                -\frac{3}{64} \zeta_2 \zeta_3^2
                +\frac{3115413}{4096} \zeta_4
                -\frac{585}{16} \zeta_2 \zeta_4
                -\frac{1067453}{3584} \zeta_3 \zeta_4
                +\frac{179855}{128} \zeta_5
                -\frac{2476517}{2240} \zeta_2 \zeta_5
\nonumber \\ &
                -\frac{3}{80} \zeta_3 \zeta_5
                +\frac{12960955}{8192} \zeta_6
                +\frac{110413}{32} \zeta_7
                -\frac{6095}{16384} \zeta_8
                -\biggl(
                        \frac{1155}{16}
                        +\frac{45}{8} \zeta_2
                        -\frac{1}{2} \zeta_3
                \biggr) B_4
\nonumber \\ &
                -\biggl(
                        1740
                        +72 \zeta_2
                \biggr) \text{Li}_5\left(\frac{1}{2}\right)
                -2160 \text{Li}_6\left(\frac{1}{2}\right)
                -1728 \text{Li}_7\left(\frac{1}{2}\right)
                -\biggl(
                        \frac{2900}{7} \zeta_3^2
                        +\frac{7395}{4} \zeta_4
\nonumber \\ &
                        +\frac{7763}{8} \zeta_6
                \biggr) \ln(2)
                -\biggl(
                        -\frac{2295}{2} \zeta_4
                        +\frac{1395}{2} \zeta_5
                \biggr) \ln(2)^2
                -\biggl(
                        145 \zeta_2
                        +321 \zeta_4
                \biggr) \ln(2)^3
\nonumber \\ &
                +45 \zeta_2 \ln(2)^4
                -\biggl(
                        -\frac{29}{2}
                        +\frac{33}{5} \zeta_2
                \biggr) \ln(2)^5
                -3 \ln(2)^6
                +\frac{12}{35} \ln(2)^7
                -900 s_6
                +\frac{2320}{7} \ln(2) s_6
\nonumber \\ &
                -\frac{2320}{7} s_{7a}
                +\frac{2720}{7} s_{7b}
        \biggr]
        +\ep^6 \biggl[
                -\frac{50743957}{49152}
                -\frac{1108525}{8192} \zeta_2
                +\frac{2129525}{2048} \zeta_3
                +\frac{129927}{1024} \zeta_2 \zeta_3
\nonumber \\ &
                -\frac{1447405}{1536} \zeta_3^2
                -\frac{2375647}{1792} \zeta_2 \zeta_3^2
                -\frac{1}{192} \zeta_3^3
                -\frac{22542751}{16384} \zeta_4
                +\frac{5655}{64} \zeta_2 \zeta_4
                +\frac{16043715}{14336} \zeta_3 \zeta_4
\nonumber \\ &
                +\frac{186147}{896} \zeta_4^2
                -\frac{7160433}{2560} \zeta_5
                +\frac{7430559}{1792} \zeta_2 \zeta_5
                +\frac{3223817}{2240} \zeta_3 \zeta_5
                -\frac{171}{2560} \zeta_4 \zeta_5
                -\frac{376334635}{98304} \zeta_6
\nonumber \\ &
                -\frac{631}{4096} \zeta_3 \zeta_6
                -\frac{11593125}{896} \zeta_7
                -\frac{9}{448} \zeta_2 \zeta_7
                -\frac{3996260561}{1376256} \zeta_8
                -\frac{1}{96} \zeta_9
                +\biggl(
                        \frac{8281}{64}
                        +\frac{435}{32} \zeta_2
\nonumber \\ &
                        -\frac{15}{8} \zeta_3
                        +\frac{14319}{448} \zeta_4
                \biggr) B_4
                +\biggl(
                        3465
                        +270 \zeta_2
                        -\frac{14088}{7} \zeta_3
                \biggr) \text{Li}_5\left(\frac{1}{2}\right)
                +6480 \text{Li}_7\left(\frac{1}{2}\right)
\nonumber \\ &
                +\biggl(
                        5220
                        +216 \zeta_2
                \biggr) \text{Li}_6\left(\frac{1}{2}\right)
                +5184 \text{Li}_8\left(\frac{1}{2}\right)
                -\biggl(
                        -\frac{10875}{7} \zeta_3^2
                        -\frac{58905}{16} \zeta_4
                        +\frac{24531}{7} \zeta_3 \zeta_4
\nonumber \\ &
                        +\frac{24480}{7} \zeta_2 \zeta_5
                        -\frac{116445}{32} \zeta_6
                \biggr) \ln(2)
                -\biggl(
                        \frac{2295}{14} \zeta_3^2
                        +\frac{22185}{8} \zeta_4
                        -\frac{20925}{8} \zeta_5
                        +\frac{2229}{16} \zeta_6
                \biggr) \ln(2)^2
\nonumber \\ &
                -\biggl(
                        -\frac{1155}{4} \zeta_2
                        +\frac{1174}{7} \zeta_2 \zeta_3
                        -\frac{4815}{4} \zeta_4
                \biggr) \ln(2)^3
                -\biggl(
                        \frac{435}{4} \zeta_2
                        +\frac{963}{4} \zeta_4
                \biggr) \ln(2)^4
                -\biggl(
                        \frac{231}{8}
\nonumber \\ &
                        -\frac{99}{4} \zeta_2
                        -\frac{587}{35} \zeta_3
                \biggr) \ln(2)^5
                -\biggl(
                        -\frac{29}{4}
                        +\frac{33}{10} \zeta_2
                \biggr) \ln(2)^6
                -\frac{9}{7} \ln(2)^7
                +\frac{9}{70} \ln(2)^8
                -\biggl(
                        -2175
\nonumber \\ &
                        +\frac{20850}{7} \zeta_2
                \biggr) s_6
                -\frac{8700}{7} \ln(2) s_6
                -\frac{4080}{7} \ln(2)^2 s_6
                +\frac{8700}{7} s_{7a}
                -\frac{10200}{7} s_{7b}
                +\frac{47969}{28} s_{8a}
\nonumber \\ &
                +\frac{75984}{7} s_{8b}
                +\frac{8160}{7} s_{8c}
                +\frac{6960}{7} s_{8d}
        \biggr]
        +\mathcal{O}(\ep^7)
        \Biggr\}.
\end{align}%

Besides the usual zeta values $\zeta_k$ and the polylogarithms evaluated at
$\frac{1}{2}$, $\mathrm{Li}_n(\frac{1}{2})$, also the following constants appear
\cite{Vermaseren:1998uu,Blumlein:2009cf}
\begin{align*}
  &\begin{aligned} \,
    B_4 &= -4 \zeta_2\ln^2(2)
           +\frac{2}{3}\ln^4(2) 
           -\frac{13}{2}\zeta_4
           +16 \mathrm{Li}_4\Bigl(\frac{1}{2}\Bigr)
    \,,
  \end{aligned} \\
  &\begin{aligned}
    s_6    &= \sigma_{-5,-1}           \,, &
    s_{7a} &= \sigma_{-5,1,1}          \,, &
    s_{7b} &= \sigma_{5,-1,-1}         \,, \\
    s_{8a} &= \sigma_{5,3}             \,, &
    s_{8b} &= \sigma_{-7,-1}           \,, &
    s_{8c} &= \sigma_{-5,-1,-1,-1}     \,, &
    s_{8d} &= \sigma_{-5,-1,1,1}       \,,
  \end{aligned}
\end{align*}
where the $\sigma_{\vec{a}}$ are multiple zeta values, defined by
$\sigma_{\vec{a}} = \lim_{N \to \infty} S_{\vec{a}}(N)$. Except for the zeta
values $\zeta_k$ and $B_4$ \cite{Broadhurst:1991fi} the remaining multiple zeta
values cancel in the final results for the operator matrix elements.

Similar results are obtained for this integral in all even dimensions $D=4+\ep$
to $D=30+\ep$ up to and including the $\mathcal{O}(\ep^6)$ terms. This allows
the calculation of moments up to $N=13$ for the master integrals with operator
insertions. Since this exceeds the number of required initial values for the
recurrences, we can use the remaining moments as a cross check for their
solutions.

\vspace*{5mm}\noindent
{\bf Acknowledgment.}~The diagrams were drawn using {\tt Axodraw}~\cite{Collins:2016aya}.
We would like to thank C.~Anastasiou, M.~Kauers, P.~Marquard, M.~Round, and Y.~Schr\"oder for discussions. 
This work was supported in 
part by the Austrian Science Fund (FWF) grants P20347-N18 and SFB F50 (F5009-N15) and the 
European Commission through contract PITN-GA-2012-316704 ({HIGGSTOOLS}).
We gratefully acknowledge support by the MITP and the Institute for Physics at Johannes 
Gutenberg University, Mainz, where part of the research by A.v.M.\ presented in this article
was performed.

\newpage


\begin{thebibliography}{99}
%
\bibitem{Accardi:2016ndt}
  A.~Accardi {\it et al.},
  Eur.\ Phys.\ J.\ C {\bf 76} (2016) no.8,  471
  [arXiv:1603.08906 [hep-ph]].
%
\bibitem{Alekhin:2017kpj}
  S.~Alekhin, J.~Bl\"umlein, S.~Moch and R.~Placakyte,
  {\it Parton Distribution Functions, $\alpha_s$ and Heavy-Quark Masses for LHC Run II},
  arXiv:1701.05838 [hep-ph].
%
\bibitem{alphas}
  S.~Bethke {\it et al.},
  {\it Workshop on Precision Measurements of $\alpha_s$}
  arXiv:1110.0016 [hep-ph];\\
  S.~Moch {\it et al.},
  {\it High precision fundamental constants at the TeV scale},
  arXiv:1405.4781 [hep-ph];\\
  S.~Alekhin, J.~Bl\"umlein and S.O.~Moch,
  Mod.\ Phys.\ Lett.\ A {\bf 31} (2016) no.25,  1630023.
%
\bibitem{FUTURE}
  D.~Boer {\it et al.},
  {\it Gluons and the quark sea at high energies: Distributions, polarization, tomography},
  arXiv:1108.1713 [nucl-th];\\
  J.~L.~Abelleira Fernandez {\it et al.} [LHeC Study Group],
  J.\ Phys.\ G {\bf 39} (2012) 075001
  [arXiv:1206.2913 [physics.acc-ph]].

%
\bibitem{Moch:2004pa}
  S.~Moch, J.A.M.~Vermaseren and A.~Vogt,
  Nucl.\ Phys.\ B {\bf 688} (2004) 101
  [hep-ph/0403192].
%
\bibitem{Vogt:2004mw}
  A.~Vogt, S.~Moch and J.A.M.~Vermaseren,
  Nucl.\ Phys.\ B {\bf 691} (2004) 129
  [hep-ph/0404111].
%
\bibitem{MOMA}
  S.A.~Larin, T.~van Ritbergen and J.A.M.~Vermaseren,
  Nucl.\ Phys.\ B {\bf 427} (1994) 41;\\
  S.A.~Larin, P.~Nogueira, T.~van Ritbergen and J.A.M.~Vermaseren,
  Nucl.\ Phys.\ B {\bf 492} (1997) 338
  [hep-ph/9605317];\\
  A.~Retey and J.A.M.~Vermaseren,
  Nucl.\ Phys.\ B {\bf 604} (2001) 281
  [hep-ph/0007294];\\
  J.~Bl\"umlein and J.A.M.~Vermaseren,
  Phys.\ Lett.\ B {\bf 606} (2005) 130
  [hep-ph/0411111];\\
  A.A.~Bagaev, A.V.~Bednyakov, A.F.~Pikelner and V.N.~Velizhanin,
  Phys.\ Lett.\ B {\bf 714} (2012) 76
  [arXiv:1206.2890 [hep-ph]];\\
  V.N.~Velizhanin,
  Nucl.\ Phys.\ B {\bf 864} (2012) 113
  [arXiv:1203.1022 [hep-ph]].
%
\bibitem{Blumlein:2009rg}
  J.~Bl\"umlein, S.~Klein and B.~T\"odtli,
  Phys.\ Rev.\ D {\bf 80} (2009) 094010
  [arXiv:0909.1547 [hep-ph]].
%
\bibitem{Bierenbaum:2009mv}
  I.~Bierenbaum, J.~Bl\"umlein and S.~Klein,
  Nucl.\ Phys.\ B {\bf 820} (2009) 417
  [arXiv:0904.3563 [hep-ph]].
%
\bibitem{Ablinger:2015tua}
  J.~Ablinger, A.~Behring, J.~Bl\"umlein, A.~De Freitas, A.~von Manteuffel and C.~Schneider,
  Comput.\ Phys.\ Commun.\  {\bf 202} (2016) 33
  [arXiv:1509.08324 [hep-ph]].
%
\bibitem{Blumlein:2017dxp}
  J.~Bl\"umlein and C.~Schneider,
  {\it The Method of Arbitrarily Large Moments to Calculate Single Scale Processes in Quantum 
  Field Theory}, Phys. Lett. B (2017) in print, arXiv:1701.04614 [hep-ph]. 
%
\bibitem{Ablinger:2014lka}
  J.~Ablinger, J.~Bl\"umlein, A.~De Freitas, A.~Hasselhuhn, A.~von Manteuffel, M.~Round, 
  C.~Schneider and F.~Wi\ss{}brock,
  Nucl.\ Phys.\ B {\bf 882} (2014) 263
  [arXiv:1402.0359 [hep-ph]].
%
\bibitem{Ablinger:2014vwa}
  J.~Ablinger, A.~Behring, J.~Bl\"umlein, A.~De Freitas, A.~Hasselhuhn, A.~von Manteuffel,
  M.~Round, C.~Schneider, and F.~Wi\ss{}brock,
  Nucl.\ Phys.\ B {\bf 886} (2014) 733
  [arXiv:1406.4654 [hep-ph]].
%
\bibitem{Ablinger:2014nga}
  J.~Ablinger, A.~Behring, J.~Bl\"umlein, A.~De Freitas, A.~von Manteuffel and C.~Schneider,
  Nucl.\ Phys.\ B {\bf 890} (2014) 48
  [arXiv:1409.1135 [hep-ph]].
%
\bibitem{TWOMASS}
J.~Ablinger, J.~Bl\"umlein, A.~De Freitas, A.~Hasselhuhn, C.~Schneider  and  
F.~Wi\ss{}brock, {\it Three Loop Massive Operator Matrix Elements and Asymptotic Wilson Coefficients
with Two Different Masses}, DESY 14--019.
%
\bibitem{Blumlein:2003gb}
  J.~Bl\"umlein,
  Comput.\ Phys.\ Commun.\  {\bf 159} (2004) 19
  [hep-ph/0311046].
%
\bibitem{Vermaseren:1998uu}
  J.A.M.~Vermaseren,
  Int.\ J.\ Mod.\ Phys.\ A {\bf 14} (1999) 2037
  [hep-ph/9806280].
%
\bibitem{Blumlein:1998if}
  J.~Bl\"umlein and S.~Kurth,
  Phys.\ Rev.\ D {\bf 60} (1999) 014018
  [hep-ph/9810241].
%
\bibitem{Remiddi:1999ew}
  E.~Remiddi and J.A.M.~Vermaseren,
  Int.\ J.\ Mod.\ Phys.\ A {\bf 15} (2000) 725
  [hep-ph/9905237].
%
\bibitem{Buza:1995ie}
  M.~Buza, Y.~Matiounine, J.~Smith, R.~Migneron and W.L.~van Neerven,
  Nucl.\ Phys.\  B {\bf 472} (1996) 611
  [arXiv:hep-ph/9601302].  
%
\bibitem{Bierenbaum:2007qe}
  I.~Bierenbaum, J.~Bl\"umlein and S.~Klein,
  Nucl.\ Phys.\  B {\bf 780} (2007) 40
  [arXiv:hep-ph/0703285].
%
\bibitem{Bierenbaum:2008yu}
  I.~Bierenbaum, J.~Bl\"umlein, S.~Klein and C.~Schneider,
  Nucl.\ Phys.\  B {\bf 803} (2008) 1
  [arXiv:0803.0273 [hep-ph]].
%
\bibitem{Buza:1996wv}
  M.~Buza, Y.~Matiounine, J.~Smith and W.L.~van Neerven,
  Eur.\ Phys.\ J.\  C {\bf 1} (1998) 301
  [arXiv:hep-ph/9612398].
%
\bibitem{Bierenbaum:2009zt}
  I.~Bierenbaum, J.~Bl\"umlein and S.~Klein,
  Phys.\ Lett.\  B {\bf 672} (2009) 401
  [arXiv:0901.0669 [hep-ph]].
%
\bibitem{Nogueira:1991ex}
  P.~Nogueira,
  J.\ Comput.\ Phys.\  {\bf 105} (1993) 279.
%
\bibitem{vanRitbergen:1998pn}
  T.~van Ritbergen, A.N.~Schellekens and J.A.M.~Vermaseren,
  Int.\ J.\ Mod.\ Phys.\ A {\bf 14} (1999) 41
  [hep-ph/9802376].
%
\bibitem{Studerus:2009ye}
  C.~Studerus,
  Comput.\ Phys.\ Commun.\  {\bf 181} (2010) 1293
  [arXiv:0912.2546 [physics.comp-ph]].
%
\bibitem{vonManteuffel:2012np}
  A.~von Manteuffel and C.~Studerus,
  {\it Reduze 2 - Distributed Feynman Integral Reduction},
  arXiv:1201.4330 [hep-ph].
%
\bibitem{FERMAT}
R.H.~Lewis, {\it Computer Algebra System {\tt Fermat}}, {\tt http://home.bway.net/lewis}.
%
\bibitem{Bauer:2000cp}
  C.W.~Bauer, A.~Frink and R.~Kreckel,
  J.\ Symb.\ Comput.\  {\bf 33} (2000) 1
  [cs/0004015 [cs-sc]].
%
\bibitem{GUESS}
M.~Kauers, {\it Guessing Handbook}, JKU Linz, Technical Report RISC 09-07.
%
\bibitem{SIG1}
C.~Schneider, {S\'em.~Lothar. Combin.\/} {\bf 56} (2007) 1,
 article B56b.
%
\bibitem{SIG2}
C.~Schneider, Simplifying Multiple Sums in Difference Fields, in:~{{\sf Computer 
Algebra in Quantum Field Theory: Integration,
  Summation and Special Functions}\/} Texts and Monographs in Symbolic
  Computation eds. C.~Schneider and J.~Bl\"umlein  (Springer, Wien, 2013) 325 
  arXiv:1304.4134 [cs.SC].
%
\bibitem{HARMONICSUMS}
  J.~Ablinger,
  PoS (LL2014) 019   [arXiv:1407.6180 [cs.SC]];\\
  J.~Ablinger,
  {\sf A Computer Algebra Toolbox for Harmonic Sums Related to Particle Physics}, Diploma Thesis, J. Kepler University Linz, 2009,
  arXiv:1011.1176 [math-ph];
%
\bibitem{Ablinger:PhDThesis}
  J.~Ablinger,
  {\sf Computer Algebra Algorithms for Special Functions in Particle Physics},
  Ph.D. Thesis, J. Kepler University Linz, 2012,
  arXiv:1305.0687 [math-ph];
%
\bibitem{Ablinger:2011te}
  J.~Ablinger, J.~Bl\"umlein and C.~Schneider,
  J.\ Math.\ Phys.\  {\bf 52} (2011) 102301
  [arXiv:1105.6063 [math-ph]].
%
\bibitem{Ablinger:2013cf}
  J.~Ablinger, J.~Bl\"umlein and C.~Schneider,
  J.\ Math.\ Phys.\  {\bf 54} (2013) 082301
  [arXiv:1302.0378 [math-ph]].
%
\bibitem{Ablinger:2014bra}
  J.~Ablinger, J.~Bl\"umlein, C.G.~Raab and C.~Schneider,
  J.\ Math.\ Phys.\  {\bf 55} (2014) 112301
  [arXiv:1407.1822 [hep-th]].
%
\bibitem{Karr:81}
M.~Karr, {J.~ACM} {\bf 28} (1981) 305.
%
\bibitem{Schneider:01}
C.~Schneider,
{\sf Symbolic Summation in Difference Fields\/} Ph.D. Thesis
RISC, Johannes Kepler University, Linz technical report 01-17 (2001).
%
\bibitem{Schneider:05a}
C. Schneider, 
Difference Equations in ${\Pi}{\Sigma}$-Extensions.
An. Univ. Timisoara Ser. Mat.-Inform. {\bf 42} (2004) 163;\\
{J. Differ. Equations Appl.\/} {\bf 11} (2005) 799
;\\
{
Appl. Algebra Engrg. Comm. Comput. {\bf16}(2005) 1.
}
%
\bibitem{Schneider:07d}
C.~Schneider, {J. Algebra Appl.\/} {\bf 6} (2007) 415.
%
\bibitem{Schneider:10b}
C.~Schneider, {\sf {Motives, Quantum Field Theory, and Pseudodifferential
  Operators}\/} ({\sf Clay Mathematics Proceedings\/} Vol.~{\bf{12}} ed. A.~Carey,
  D.~Ellwood, S.~Paycha and S.~Rosenberg,(Amer. Math. Soc) (2010), 285 
  arXiv:0904.2323.
%
\bibitem{Schneider:10c}
C.~Schneider, {Ann. Comb.\/} {\bf 14} (2010)  533
[arXiv:0808.2596].
%
\bibitem{Schneider:15a}
C.~Schneider,
in: Computer Algebra and Polynomials, Applications of Algebra and Number Theory, 
J.~Gutierrez, J.~Schicho, M.~Weimann (ed.), Lecture Notes in Computer Science 
(LNCS) 8942 (2015), 157
[arXiv:13077887 [cs.SC]].
%
\bibitem{Schneider:08c}
C.~Schneider, {J. Symbolic Comput.} {\bf 43} (2008) 611,
  [arXiv:0808.2543v1];\\
J. Symb. Comput. {\bf 72} (2016) 82,
doi:10.1016/j.jsc.2015.02.002, arXiv:1408.2776 [cs.SC].
{J. Symb. Comput. {\bf 80} (2017) 616, 
arXiv:1603.04285 [cs.SC].
}
%
\bibitem{SAGE}
Sage {\tt http://www.sagemath.org/}
%
\bibitem{GSAGE}
M.~Kauers, M.~Jaroschek, and F.~Johansson, in:
{\sf Computer Algebra and Polynomials}, 
Editors: J.~Gutierrez, J.~Schicho, Josef, M.~Weimann, Eds..
Lecture Notes in Computer Science {\bf 8942} (Springer, Berlin, 2015) 105 
[arXiv:1306.4263 [cs.SC]].
%
\bibitem{Blumlein:2009tj}
  J.~Bl\"umlein, M.~Kauers, S.~Klein and C.~Schneider,
  Comput.\ Phys.\ Commun.\  {\bf 180} (2009) 2143
  [arXiv:0902.4091 [hep-ph]].
%
\bibitem{Gross:1974cs}
  D.J.~Gross and F.~Wilczek,
  {Phys.\ Rev.}\ D {\bf 9} (1974) 980.
%
\bibitem{Georgi:1951sr}
  H.~Georgi and H.D.~Politzer,
  {Phys.\ Rev.}\ D {\bf 9} (1974) 416.
%
\bibitem{Floratos:1978ny}
E.G.~Floratos, D.A.~Ross and C.T.~Sachrajda,
  {Nucl.\ Phys.}\ B {\bf 152} (1979) 493.
%
\bibitem{GonzalezArroyo:1979he}
  A.~Gonzalez-Arroyo and C.~Lopez,
  {Nucl.\ Phys.}\ B {\bf 166} (1980) 429.
%
\bibitem{Furmanski:1980cm}
  W.~Furmanski and R.~Petronzio,
  {Phys.\ Lett.}\ B {\bf 97} (1980) 437.
%
\bibitem{Hamberg:1991qt}
  R.~Hamberg and W.L.~van Neerven,
  {Nucl.\ Phys.}\ B {\bf 379} (1992) 143.
%
\bibitem{Ellis:1996nn}
  R.K.~Ellis and W.~Vogelsang,
  hep-ph/9602356.
%
\bibitem{Moch:1999eb}
  S.~Moch and J.A.M.~Vermaseren,
  {Nucl.\ Phys.}\ B {\bf 573} (2000) 853
  [hep-ph/9912355].
%
\bibitem{Ablinger:2010ty}
  J.~Ablinger, J.~Bl\"umlein, S.~Klein, C.~Schneider and F.~Wi\ss{}brock,
  Nucl.\ Phys.\ B {\bf 844} (2011) 26
  [arXiv:1008.3347 [hep-ph]].
%
\bibitem{HYP}
F.~Klein, {\sf Vorlesungen \"uber die hypergeometrische Funktion}, Wintersemester 1893/94,
Die Grundlehren der Mathematischen Wissenschaften {\bf 39}, (Springer, Berlin, 1933);
\\
W.N. Bailey, {\sf Generalized Hypergeometric Series}, (Cambridge University
Press,  Cambridge, 1935);
\\
P. Appell and J. Kamp\'{e} de F\'{e}riet, {\sf Fonctions
Hyperg\'{e}om\'{e}triques et Hypersp\'{e}riques, Polynomes D' Hermite},
(Gauthier-Villars, Paris, 1926);
\\
P. Appell, {\sf Les Fonctions Hyperg\"{e}om\'{e}triques de Plusieur
Variables}, (Gauthier-Villars, Paris, 1925);
\\
J. Kamp\'{e} de F\'{e}riet, {\sf La fonction
hyperg\"{e}om\'{e}trique},(Gauthier-Villars, Paris, 1937);
\\
H. Exton, {\sf Multiple Hypergeometric Functions and Applications},
(Ellis Horwood, Chichester, 1976);\\
H. Exton, {\sf Handbook of Hypergeometric Integrals},
(Ellis Horwood, Chichester, 1978);
\\
H.M. Srivastava and P.W. Karlsson, {\sf Multiple Gaussian Hypergeometric
Series}, (Ellis Horwood, Chicester, 1985);\\
  M.J.~Schlosser, in: {\sf Computer Algebra in Quantum Field Theory: Integration, Summation and
  Special Functions}, C. Schneider, J. Bl\"umlein, Eds.,~p.~305, (Springer, Wien, 2013)
  [arXiv:1305.1966 [math.CA]].
%
\bibitem{SLATER}
L.J.~Slater, {\sf Generalized hypergeometric functions}, (Cambridge, Cambridge University Press, 
1966).
%
\bibitem{MB1}
E.W. Barnes, Proc. Lond. Math. Soc. (2) {\bf 6} (1908) 141; Quart.
Journ. Math. {\bf 41} (1910) 136;\\ 
H. Mellin,
Math. Ann. {\bf 68} (1910) 305.
%
\bibitem{MB}
  M.~Czakon,
  Comput.\ Phys.\ Commun.\  {\bf 175} (2006) 559
  [hep-ph/0511200];\\
  A.V.~Smirnov and V.A.~Smirnov,
  Eur.\ Phys.\ J.\ C {\bf 62} (2009) 445
  [arXiv:0901.0386 [hep-ph]].
%
\bibitem{EMSSP}
  J.~Ablinger, J.~Bl\"umlein, S.~Klein and C.~Schneider,
  Nucl.\ Phys.\ Proc.\ Suppl.\  {\bf 205-206} (2010) 110
  [arXiv:1006.4797 [math-ph]];\\
  J.~Bl\"umlein, A.~Hasselhuhn and C.~Schneider,
  PoS (RADCOR 2011) 032
  [arXiv:1202.4303 [math-ph]];\\
  C. Schneider,
  Computer Algebra Rundbrief {\bf 53} (2013) 8
  ;\\
  C.~Schneider,
  J.\ Phys.\ Conf.\ Ser.\  {\bf 523} (2014) 012037
  [arXiv:1310.0160 [cs.SC]].
%
\bibitem{DEQ}
  A.V.~Kotikov,
  Phys.\ Lett.\ B {\bf 254} (1991) 158;\\
  M.~Caffo, H.~Czyz, S.~Laporta and E.~Remiddi,
  Acta Phys.\ Polon.\ B {\bf 29} (1998) 2627
  [hep-th/9807119];
  Nuovo Cim.\ A {\bf 111} (1998) 365
  [hep-th/9805118];\\
  T.~Gehrmann and E.~Remiddi,
  Nucl.\ Phys.\ B {\bf 580} (2000) 485
  [hep-ph/9912329];\\
  A.V. Kotikov, In : Subtleties in quantum field theory, ed. D. Diakonov, 
  150 
  [arXiv:1005.5029 [hep-th]];
  Theor. Math. Phys. 176 (2013) 913 [arXiv:1212.3732 [hep-ph]]; 
  Phys. Part. Nucl. 44 (2013) 374;\\
  J.M.~Henn,
  Phys.\ Rev.\ Lett.\  {\bf 110} (2013) 25,  251601
  [arXiv:1304.1806 [hep-th]].
%
\bibitem{Gerhold:2002}
S.~Gerhold, {\sf Uncoupling systems of linear {O}re operator equations},
Master's thesis, RISC, J.~Kepler University, Linz, 2002;
The package {\tt OreSys} has been tuned for our spefic calculations by
the last author.
%
\bibitem{AZ}
G.~Almkvist and D.~Zeilberger, J. Symb. Comp. {\bf 10} (1990) 571; 
\\
M.~Apagodu and D.~Zeilberger,
Adv. Appl. Math. (Special Regev Issue), {\bf 37} (2006) 139. 
%
\bibitem{Ablinger:2014uka}
J.~Ablinger, J.~Bl{\"u}mlein, A.~De~Freitas, A.~Hasselhuhn, A.~von Manteuffel,
M.~Round, and C.~Schneider, 
{Nucl. Phys.} {\bf 885} (2014) 280  
[arXiv:1405.4259 [hep-ph]].
%
\bibitem{Blumlein:2012vq}
  J.~Bl\"umlein, A.~Hasselhuhn, S.~Klein and C.~Schneider,
  Nucl.\ Phys.\ B {\bf 866} (2013) 196
  [arXiv:1205.4184 [hep-ph]].
%
\bibitem{Bennett:1997ch}
  J.F.~Bennett and J.A.~Gracey,
  Nucl.\ Phys.\ B {\bf 517} (1998) 241
  [hep-ph/9710364].
%
\bibitem{Brown:2008um}
  F.~Brown,
  Commun.\ Math.\ Phys.\  {\bf 287} (2009) 925
  [arXiv:0804.1660 [math.AG]].
%
\bibitem{Ablinger:2014yaa}
  J.~Ablinger, J.~Bl\"umlein, C.~Raab, C.~Schneider and F.~Wi\ss{}brock,
  Nucl.\ Phys.\ B {\bf 885} (2014) 409
  [arXiv:1403.1137 [hep-ph]].
%
\bibitem{Panzer:2014caa}
  E.~Panzer,
  Comput.\ Phys.\ Commun.\  {\bf 188} (2015) 148
  [arXiv:1403.3385 [hep-th]].
%
\bibitem{Vermaseren:2000nd}
  J.A.M.~Vermaseren,
  {\it New features of FORM},
  math-ph/0010025.
%
\bibitem{Steinhauser:2000ry}
M.~Steinhauser, 
Comput. Phys. Commun. {\bf 134} (2001) 335, 
[arXiv:hep-ph/0009029].
%
\bibitem{Klein:2009ig}
S.W.G. Klein, 
{\sf Mellin Moments of Heavy Flavor Contributions to $F_2(x,Q^2)$ at NNLO},
\newblock PhD thesis, TU Dortmund, 2009.
%
\bibitem{Itzykson:1980rh}
C.~Itzykson and J.B.~Zuber, {\sf {Quantum Field Theory}}.
\newblock International Series In Pure and Applied Physics. (McGraw-Hill, New
York, 1980).
%
\bibitem{Smirnov:2006ry}
V.A.~Smirnov, {\sf {Feynman integral calculus}}.
\newblock (Springer, Berlin, 2006).
%
\bibitem{Bogner:2010kv}
C.~Bogner and S.~Weinzierl, 
{Int. J. Mod. Phys.} A {\bf 25} (2010) 2585 
[arXiv:1002.3458 [hep-ph]].
%
\bibitem{Tarasov:1996br}
O.V.~Tarasov, 
{Phys. Rev.} D {\bf 54} (1996) 6479. 
[arXiv:hep-th/9606018 [hep-th]]
%
\bibitem{Lee:2009dh}
R.N.~Lee, 
Nucl. Phys. B {\bf 830} (2010) 474   
[arXiv:0911.0252 [hep-ph]] 
%
\bibitem{IBP}
J. Lagrange, {\sf Nouvelles recherches sur la nature et la propagation
du son}, Miscellanea Taurinensis, t. II, 1760-61; Oeuvres t. I, p. 263;\\
C.F. Gau\ss{}, {Theoria attractionis corporum sphaeroidicorum ellipticorum
homogeneorum methodo novo tractate}, Commentationes societas scientiarum
Gottingensis recentiores, Vol III, 1813, Werke Bd. {\bf V} pp. 5-7;\\
G. Green, {\sf Essay on the Mathematical Theory of Electricity and
Magnetism}, Nottingham, 1828 [Green Papers, pp. 1-115];\\
M. Ostrogradski, Mem. Ac. Sci. St. Peters., {\bf 6}, (1831) 39;\\
  K.G.~Chetyrkin and F.V.~Tkachov,
  Nucl.\ Phys.\ B {\bf 192} (1981) 159.
%
\bibitem{Broadhurst:1991fi}
D.J.~Broadhurst, 
Z. Phys. C {\bf 54} (1992) 599.
%
\bibitem{Bekavac:2009gz}
  S.~Bekavac, A.~G.~Grozin, D.~Seidel and V.~A.~Smirnov,
  Nucl.\ Phys.\ B {\bf 819} (2009) 183
  [arXiv:0903.4760 [hep-ph]].
%
\bibitem{Grigo:2012ji}
J.~Grigo, J.~Hoff, P.~Marquard, and M.~Steinhauser, 
Nucl. Phys. B {\bf864} (2012) 580   
[arXiv:1206.3418 [hep-ph]].
%
\bibitem{Schroeder:LL2016}
Y.~Schr{\"o}der, 
{\it Five loop massive tadpoles}, Presentation at Loops
and Legs in QFT, 2016,
\newblock
{\tt https://indico.desy.de/conferenceOtherViews.py?view=standard\&confId=12010};\\
  T.~Luthe and Y.~Schr\"oder,
  PoS (LL2016) 074
  [arXiv:1609.06786 [hep-ph]].
%
\bibitem{UNCOUPL}
A.~Danilevski\u{\i},
Mat. Sbornik {\bf 2} (1937) 169;\\ 
M.A. Barkatou,
Appl. Algebra Engrg. Comm. Comput. {\bf 4}(3) (1993) 185;\\ 
B.~Z\"urcher, {\sf Rationale Normalformen von pseudo-linearen Abbildungen},
Master's thesis, ETH Z\"urich (1994); \\
M. Bronstein and M. Petkov\v{s}ek,
Theor. Comput. Sci., {\bf 157} (1) (1996) 3;\\ 
S.A.~Abramov and E.V.~Zima,
Proc. Int. Conf. on Computational Modelling and Computing in Physics, Dubna, RU, Sept. 16-26 (1996) 
16;\\ 
A.~Bostan, F.~Chyzak, E.~de~Panafieu,
ISSAC 2013, Boston, arXiv:1301.5414 [cs.SC] and references therein.
%
\bibitem{KAMKE}
E.~Kamke, {\sf Differentialgleichungen: L\"osungsmethoden und L\"osungen}, (Geest \& Portig, Leipzig, 1967), 
8th~Edition. 
%
\bibitem{Blumlein:2009cf}
J.~Bl{\"u}mlein, D.J.~Broadhurst, and J.A.M.~Vermaseren, 
Comput. Phys. Commun. {\bf 181} (2010) 582    
[arXiv:0907.2557 [math-ph]].
%
\bibitem{Collins:2016aya}
  J.C.~Collins and J.A.M.~Vermaseren,
  {\it Axodraw Version 2},
  arXiv:1606.01177 [cs.OH].
\end{thebibliography}
\end{document}